\documentclass[a4paper,11pt]{article}
\usepackage{jheppub}
\usepackage{xcolor}
\usepackage{physics}
\usepackage{xspace}
\usepackage{mathrsfs}
\usepackage{soul}
\usepackage{mathtools}
\usepackage{tensor}
\usepackage{tabularx}
\usepackage{cancel}
\usepackage{booktabs}
\usepackage{marginnote}
\usepackage{subfigure}
\usepackage{diagbox}
\usepackage{graphicx}
\usepackage{slashed}
\allowdisplaybreaks[4]

\makeatletter
\g@addto@macro\bfseries{\boldmath}
\makeatother

\title{Scalar Triple-Heavy Tetraquark States With Quark Content $cc\bar{c}\bar{s}$}

\author{Yong-Jiang Xu$^{a}$, Luo-Geng Tian$^{a}$, Yuan Li$^{b}$, Jin-Yun Wu$^b$, Yong-Lu Liu$^{c}$, and Ming-Qiu Huang$^{c}$}

\affiliation[a]{Test and Training Base, National University of Defense Technology, Xi$^{\:\prime}$an 710106, China}
\affiliation[b]{College of Information and Communication, Wuhan 430030, China}
\affiliation[c]{Department of Physics, College of Science, National University of Defense Technology, Changsha 410073, China}

\emailAdd{xuyongjiang@nudt.edu.cn}
\emailAdd{yongluliu@nudt.edu.cn}
\emailAdd{mqhuang@nudt.edu.cn}

\abstract{
In this paper, we study the scalar triple-heavy tetraquark states with quark content $cc\bar{c}\bar{s}$, $\eta_{c}D_{s}$ molecular state and $[cc]_{A(T)}[\bar{c}\bar{s}]_{A(T)}$ compact tetraquark states, by the QCD sum rule method. First, we construct the needed interpolating currents, $J^{M}(x)$, $J^{T_{1}}(x)$, and $J^{T_{2}}(x)$. Then, we derive the sum rules for the masses and the current coupling constants. Finally, we numerically analyze these sum rules, and find $m_{M}=4.9392^{+0.0851}_{-0.0817}~\mbox{GeV}$, and $\lambda_{M}=2.8857^{+0.5729}_{-0.4928}\times10^{-2}~\mbox{GeV}^{5}$ for the mass and the current coupling constant of the $\eta_{c}D_{s}$ molecular state, $m_{T_{1}}=5.0774^{+0.0708}_{-0.0641}~\mbox{GeV}$, and $\lambda_{T_{1}}=1.0436^{+0.1862}_{-0.1573}\times10^{-1}~\mbox{GeV}^{5}$ for the mass and the current coupling constant of the $[cc]_{A}[\bar{c}\bar{s}]_{A}$ compact tetraquark state, $m_{T_{2}}=5.0679^{+0.0839}_{-0.0721}~\mbox{GeV}$, and $\lambda_{T_{2}}=2.0316^{+0.4119}_{-0.3119}\times10^{-1}~\mbox{GeV}^{5}$ for the mass and the current coupling constant of the $[cc]_{T}[\bar{c}\bar{s}]_{T}$ compact tetraquark state.
}

\begin{document}
\maketitle
\flushbottom

\section{Introduction}\label{sec1}

Heavy-flavor hadrons have been one of the most important research fields in fundamental particle physics since the charm-quark and bottom-quark were discovered experimentally. There have been plenty of works on their properties, both theoretical and experimental\cite{ParticleDataGroup:2024cfk}. In particular, since the observation of $X(3872)$ in 2003\cite{Belle:2003nnu}, there have been plenty of $X,Y,Z$ states and pentaquark states with one, two, and four heavy quarks, the so-called exotic hadron states, being reported experimentally\cite{Liu:2013waa,Hosaka:2016pey,Chen:2016qju,Richard:2016eis,Lebed:2016hpi,Guo:2017jvc,Liu:2019zoy,Brambilla:2019esw,Chen:2022asf,Wang:2025sic}. Meantime, based on different inner quark configurations, a lot of theoretical investigations have being devoted to explaining their properties, mass, decay width, production, multipole moment and so on\cite{Liu:2013waa,Hosaka:2016pey,Chen:2016qju,Richard:2016eis,Lebed:2016hpi,Guo:2017jvc,Liu:2019zoy,Brambilla:2019esw,Chen:2022asf,Wang:2025sic}. 

For the triple-heavy hadrons, although they have not been discovered experimentally until now, theoretical community has paid attention to the triple-baryons, triple-tetraquarks and triple-pentaquarks\cite{GomshiNobary:2003sf,GomshiNobary:2004mq,Brambilla:2005yk,GomshiNobary:2005ur,Jia:2006gw,GomshiNobary:2007ofo,Martynenko:2007je,Patel:2008mv,Meinel:2010pw,Chen:2011mb,Flynn:2011gf,Llanes-Estrada:2011gwu,Wang:2011ae,Albertus:2012isp,Meinel:2012qz,Aliev:2012tt,Padmanath:2013zfa,Aliev:2014lxa,Wei:2015gsa,Wei:2016jyk,Shah:2017jkr,Wang:2018utj,Shah:2018div,Yang:2019lsg,Wang:2019gal,Liu:2019vtx,Alomayrah:2020qyw,Wang:2020avt,Wu:2021tzo,Mutuk:2021zes,Huang:2021jxt,Faustov:2021qqf,Wang:2022ias,Zhao:2022vfr,Li:2022vbc,Wu:2022fpj,Zhao:2023imq,Oudichhya:2023pkg,Zhao:2023qww,Najjar:2024deh,deArenaza:2024dhe,Xie:2024lfo,Najjar:2024ngm,Dhindsa:2024erk,Yu:2025gdg,Salehi:2025hjn,Najjar:2025dzl,Chen:2016ont,Jiang:2017tdc,Liu:2019mxw,Xing:2019wil,Weng:2021ngd,Lu:2021kut,Liu:2022jdl,Mutuk:2023yev,Zhu:2023lbx,Yang:2024nyc,Zhang:2024jvv,Li:2025fmf,Galkin:2025ubt,Guo:2013xga,Chen:2017jjn,Wang:2018ihk,Li:2018vhp,Wang:2019aoc,An:2019idk,Wang:2024yjp}. Ref.\cite{Chen:2016ont} systematically investigated the mass splittings of the $QQ\bar{Q}\bar{q}$ tetraquark states and estimated their masses in the framework of the color-magnetic interaction. With the QCD sum rule method, the authors of refs.\cite{Jiang:2017tdc,Zhang:2024jvv} computed the mass spectra of the $QQ\bar{Q}\bar{q}$ tetraquark states with quantum numbers $J^{P}=0^{+}$ and $J^{P}=1^{+}$. In\cite{Weng:2021ngd}, the authors studied the mass spectrum of the $S$-wave $qQ\bar{Q}\bar{Q}$ tetraquarks in the framework of extended chromomagnetic model. In the framework of the chiral quark model with the resonating ground method, the triply heavy tetraquark states $QQ\bar{Q}\bar{q}$ with all possible quantum numbers were systematically investigated\cite{Liu:2022jdl}. Halil Mutuk studied the $S$-wave mass spectra of flavor exotic triply-heavy tetraquark states in a nonrelativistic quark model with a color interaction described by a potential computed in AdS/QCD\cite{Mutuk:2023yev}. In the present work, we focus on the triple-heavy tetraquarks with strangeness.

According to the Standard Model(SM), the strong interaction is described by the Quantum Chromodynamics(QCD), whose coupling constant runs with the energy in process. The running coupling constant leads to two important properties, asymptotic freedom at high energy and quark confinement at low energy. Due to the quark confinement, one should consider nonperturbative effects and develope nonperturbative methods as studying processes at hadron level. In 1979, Mikhail A. Shifman, A. I. Vainshtein, and Valentin I. Zakharov proposed an nonperturbative method, the QCD sum rule\cite{Shifman:1978bx,Shifman:1978by}. The QCD sum rule method is an analytic formalism firmly entrenched in QCD with minimal modeling and has been successfully applied in almost every aspect of strong interaction physics\cite{Reinders:1984sr,Colangelo:2000dp,Albuquerque:2018jkn,Wang:2025sic}. In this paper, we study the scalar triple-heavy tetraquark states with quark content $cc\bar{c}\bar{s}$, $\eta_{c}D_{s}$ molecular state, $[cc]_{A(T)}[\bar{c}\bar{s}]_{A(T)}$ compact tetraquark states, by the QCD sum rule method.

The rest of the paper is organized as follows. In Sec.\ref{sec2}, the relevant sum rules are derived. Section~\ref{sec3} is devoted to the numerical analysis, and a short summary is given in Sec.\ref{sec4}. In Appendix \ref{spectral_density}, the spectral densities are shown.
\section{The derivation of the sum rules}\label{sec2}

As the first step of the QCD sum rule method, we construct the interpolating currents with quark content $cc\bar{c}\bar{s}$ and quantum number $J^{P}=0^{+}$. For the $\eta_{c}D_{s}$ molecular state, the interpolating current is 
\begin{equation}
	J^{M}(x)=[\bar{c}_{i}(x)i\gamma_{5}c_{i}(x)][\bar{s}_{j}(x)i\gamma_{5}c_{j}(x)],
\end{equation}
where $c$ is the charm quark field, $s$ is the strange quark field, and $i,j$ are color indexes.
For the diquark-antidiquark configurations, we use the interpolating currents of the following type,
\begin{equation}
	J(x)=\epsilon_{ijn}\epsilon_{kln}[c^{T}_{i}(x)C\Gamma c_{j}(x)][\bar{c}_{k}C\Gamma\bar{s}^{T}_{l}(x)],
\end{equation}
where $c$ is the charm quark field, $s$ is the strange quark field, $C$ is the charge operator, $T$ stands for matrix transpose on the spinor index, and $i,j,k,l,n$ are color indexes. $\Gamma$ may be $1$, $\gamma_{5}$, $\gamma_{\mu}$, $\gamma_{5}\gamma_{\mu}$, or $\sigma_{\mu\nu}$. Because $\epsilon_{ijn}[c^{T}_{i}(x)C\Gamma c_{j}(x)]=\epsilon_{ijn}[c^{T}_{i}(x)(C\Gamma)^{T}c_{j}(x)]$, $C\Gamma$ must be symmetrical. Therefore, we take $\Gamma$ to be $\gamma_{\mu}$ or $\sigma_{\mu\nu}$, and the interpolating currents are
\begin{center}
	\begin{eqnarray}
		J^{T_{1}}(x)=\epsilon_{ijn}\epsilon_{kln}[c^{T}_{i}(x)C\gamma_{\mu}c_{j}(x)][\bar{c}_{k}C\gamma^{\mu}\bar{s}^{T}_{l}(x)],\nonumber\\J^{T_{2}}(x)=\epsilon_{ijn}\epsilon_{kln}[c^{T}_{i}(x)C\sigma_{\mu\nu}c_{j}(x)][\bar{c}_{k}C\sigma^{\mu\nu}\bar{s}^{T}_{l}(x)].
	\end{eqnarray}
\end{center}

We next turn to the two-point correlation functions of the above interpolating currents in the QCD vacuum:
\begin{equation}\label{correlation_function}
	\Pi(p)=i\int dx^{4}e^{ipx}\langle0\mid\mbox{T}[J(x)J^{\dagger}(0)]\mid0\rangle,
\end{equation}
where $\mbox{T}$ is the time-ordered operator, and $J(x)=J^{M}(x),J^{T_{1}}(x),J^{T_{2}}(x)$ is the interpolating current. The correlation function can be computed at both the hadron level in terms of hadronic parameters and the quark-gluon level by the operator product expansion(OPE), called physical side and theoretical side, respectively. 

In order to obtain the physical side, it is necessary to insert into the correlation function (\ref{correlation_function}) a complete set of hadrons with the same quantum numbers as the ones we are considering. Using the dispersion relation and isolating the ground term from the others, we have
\begin{equation}\label{physical_side}
	\Pi(p)=\frac{\lambda^{2}}{m^{2}-p^{2}}+\int^{\infty}_{s_{min}}ds\frac{\rho_{phys}(s)}{s-p^{2}}+\mbox{Subtraction Terms},
\end{equation}
where $\lambda$ is the pole residue defined by $\langle0\mid J(0)\mid X(p)\rangle=\lambda$ with $X(p)$ being the hadrons considered in this paper, $m$ is the hadronic mass, $s_{min}\equiv(3m_{c}+m_{s})^{2}$ with $m_{c}$ being the mass of the charm quark and $m_{s}$ the light quark's mass, and $\rho_{phys}(s)$ is the spectrum density of the higher resonances and continuum.

When we come to the theoretical side, we should substitute the interpolating currents in (\ref{correlation_function}) and contract relevant quark fields. Taking as an example the correlation function of $J^{M}(x)$, one has
\begin{eqnarray}
	\Pi(p)=i\int d^{4}xe^{ipx}&&\{Tr[(i\gamma_{5})S^{(c)}_{ik}(x)(i\gamma_{5})S^{(c)}_{ki}(-x)]Tr[(i\gamma_{5})S^{(c)}_{jl}(x)(i\gamma_{5})S^{(s)}_{lj}(-x)]\nonumber\\&&-Tr[(i\gamma_{5})S^{(c)}_{il}(x)(i\gamma_{5})S^{(s)}_{lj}(-x)(i\gamma_{5})S^{(c)}_{jk}(x)(i\gamma_{5})S^{(c)}_{ki}(-x)]\},
\end{eqnarray}
where $S^{(s)}_{ij}(x)$ and $S^{(c)}_{ij}(x)$ are the full propagators of the strange quark and the charm quark, respectively. Using the following expressions of $S^{(s)}_{ij}(x)$ and $S^{(c)}_{ij}(x)$,
\begin{eqnarray}
	S^{(s)}_{ij}(x)=&&\frac{i \not\!{x}}{2\pi^{2}x^4}\delta_{ij}-\frac{m_{s}}{4\pi^2x^2}\delta_{ij}-\frac{\langle\bar{s}s\rangle}{12}\delta_{ij}
	+i\frac{\langle\bar{s}s\rangle}{48}m_{s}\not\!{x}\delta_{ij}-\frac{x^2}{192}\langle g_{s}\bar{s}\sigma Gs\rangle \delta_{ij}\nonumber\\
	&&+i\frac{x^2\not\!{x}}{1152}m_{s}\langle g_{s}\bar{s}\sigma Gs\rangle \delta_{ij}-i\frac{g_{s}t^{a}_{ij}G^{a}_{\mu\nu}}{32\pi^2x^2}(\not\!{x}\sigma^{\mu\nu}+\sigma^{\mu\nu}\not\!{x})+\cdots,
\end{eqnarray}
\begin{eqnarray}
	S^{(c)}_{ij}(x)=i\int\frac{d^{4}k}{(2\pi)^4}e^{-ikx}&&[\frac{\not\!{k}+m_{c}}{k^2-m^{2}_{c}}\delta_{ij}
	-\frac{g_{s}t^{a}_{ij}G^{a}_{\mu\nu}}{4}\frac{\sigma^{\mu\nu}(\not\!{k}+m_{c})+(\not\!{k}+m_{c})\sigma^{\mu\nu}}
	{(k^2-m^{2}_{c})^{2}}\nonumber\\
	&&+\frac{\langle g^{2}_{s}GG\rangle}{12}\delta_{ij}m_{c}\frac{k^2+m_{c}\not\!{k}}{(k^2-m^{2}_{c})^{4}}+\cdots],
\end{eqnarray}
where $t^{a}=\frac{\lambda^{a}}{2}$ with $\lambda^{a}$ being the Gell-Mann matrix, $g_{s}$ is the strong interaction coupling constant, we obtain
\begin{equation}\label{theoratical_side}
	\Pi(p)=\int^{\infty}_{s_{min}}ds\frac{\rho_{OPE}(s)}{s-p^{2}},
\end{equation}
where $\rho_{OPE}(s)=\frac{1}{\pi}\mbox{Im}\Pi(s)$ is the theoretical spectral density given in appendix \ref{spectral_density}.

Matching the physical expression (\ref{physical_side}) with the theoretical one (\ref{theoratical_side}), one has
\begin{equation}
	\frac{\lambda^{2}}{m^{2}-p^{2}}+\int^{\infty}_{s_{min}}ds\frac{\rho_{phys}(s)}{s-p^{2}}+\mbox{Subtraction Terms}=\int^{\infty}_{s_{min}}ds\frac{\rho_{OPE}(s)}{s-p^{2}}.
\end{equation}
In order to eliminate the subtraction terms, suppress the contributions of the higher resonances and continuum, and improve the convergence of the OPE series, it is necessary to introduce the Borel transformation,
\begin{equation}
	\mathcal{B}[f(p^{2})]=\lim_{\substack{-p^{2},n\rightarrow\infty\\-p^{2}/n=M^{2}_{B}}}\frac{(-p^{2})^{n+1}}{n!}(\frac{d}{dp^{2}})f(p^{2}),
\end{equation}
where $M^{2}_{B}$ is the Borel parameter. After doing the Borel transformation, we have
\begin{equation}
	\lambda^{2}e^{-m^{2}/M^{2}_{B}}+\int^{\infty}_{s_{min}}ds\rho_{phys}(s)e^{-s/M^{2}_{B}}=\int^{\infty}_{s_{min}}ds\rho_{OPE}(s)e^{-s/M^{2}_{B}}.
\end{equation}
The spectral densities $\rho_{phys}(s)$ and $\rho_{OPE}(s)$ can be related by the quark-hadron duality,
\begin{equation}
	\rho_{phys}(s)=\rho_{OPE}(s)\theta(s-s_{0})
\end{equation}
with $s_{0}$ being the effective threshold. Therefore, we obtain
\begin{equation}\label{matching_result}
	\lambda^{2}e^{-m^{2}/M^{2}_{B}}=\int^{s_{0}}_{s_{min}}ds\rho_{OPE}(s)e^{-s/M^{2}_{B}}.
\end{equation}

Differentiating (\ref{matching_result}) with respect to $-\frac{1}{M^{2}_{B}}$ and taking the ratio of the result to (\ref{matching_result}), we get the mass sum rule
\begin{equation}
	m^{2}=\frac{\frac{d}{d(-\frac{1}{M^{2}_{B}})}\int^{s_{0}}_{s_{min}}ds\rho_{OPE}(s)e^{-s/M^{2}_{B}}}{\int^{s_{0}}_{s_{min}}ds\rho_{OPE}(s)e^{-s/M^{2}_{B}}}.
\end{equation}

\section{Numerical analysis}\label{sec3}

The sum rules contain some parameters, such as quark masses, vacuum condensates, and two auxiliary parameters. We take $m_{c}=(1.273\pm0.0046)~\mbox{GeV}$ and $m_{s}=(0.0935\pm0.0008)~\mbox{GeV}$\cite{ParticleDataGroup:2024cfk}. The values of the vacuum condensates are $\langle\bar{u}u\rangle=\langle\bar{d}d\rangle=-(0.24\pm0.01)^{3}~\mbox{GeV}^{3}$, $\langle\bar{s}s\rangle=(0.8\pm0.1)\langle\bar{u}u\rangle$, $\langle g_{s}\bar{s}\sigma Gs\rangle=(0.8\pm0.1)\langle\bar{s}s\rangle~\mbox{GeV}^{2}$, $\langle g^{2}_{s}GG\rangle=0.47~\mbox{GeV}^{4}$\cite{Shifman:1978bx,Shifman:1978by,Reinders:1984sr,Colangelo:2000dp}.

For the threshold parameter $s_{0}$ and the Borel mass $M^{2}_{B}$, we should determine their working intervals in which the mass and the pole residue vary weakly. The continuum threshold is related to the square of the first excited states having the same quantum number as the interpolating field, while the Borel parameter is determined by demanding that both the contributions of the higher states and continuum are sufficiently suppressed and the contributions coming from higher-dimensional operators are small. Specifically, we require that the pole contribution is greater than 40 percent of the total contribution, the perturbative term is larger than the quark condensate, and besides the quark condensate, the biggest condensate is less than 30 percent of the perturbation.

We define the following quantities, the ratio of the pole contribution to the total contribution (RP) and the ratio of the various condensates in the OPE series to the perturbative term ($\mbox{RD}_{i}$),
\begin{eqnarray}\label{ratio}
	&&RP\equiv\frac{\int^{s_{0}}_{s_{min}}ds\rho(s)e^{-s/M^{2}_{B}}}{\int^{\infty}_{s_{min}}ds\rho(s)e^{-s/M^{2}_{B}}},
	\nonumber\\&&RD_{i}\equiv\frac{\int^{s_{0}}_{s_{min}}ds\rho_{i}(s)e^{-s/M^{2}_{B}}}{\int^{s_{0}}_{s_{min}}ds\rho_{0}(s)e^{-\frac{s}{M^{2}_{B}}}},
\end{eqnarray}
where $i=3,4,5,7$.

\begin{figure}[htb]
	\subfigure[]{
		\includegraphics[width=4.8cm]{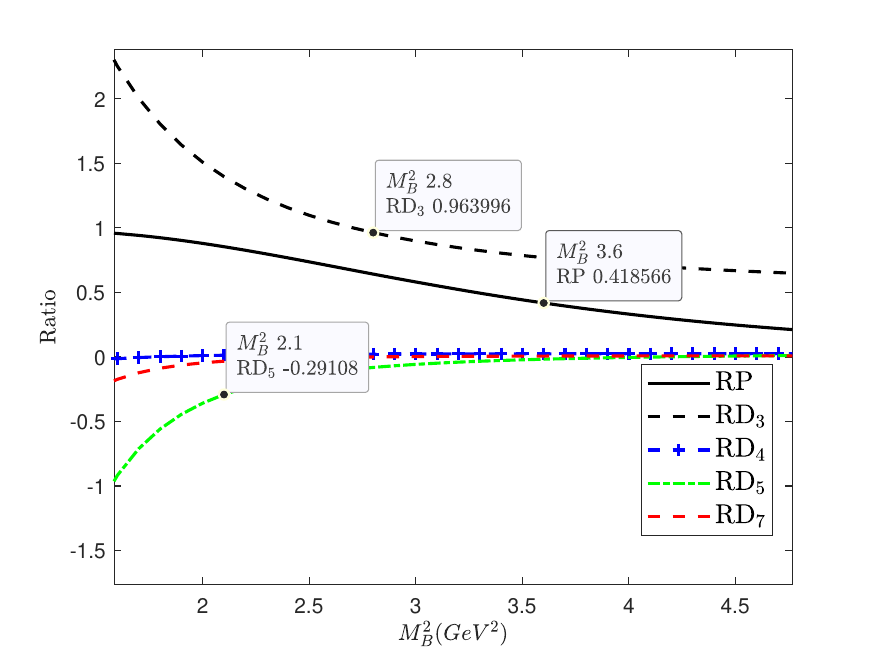}}
	\subfigure[]{
		\includegraphics[width=4.8cm]{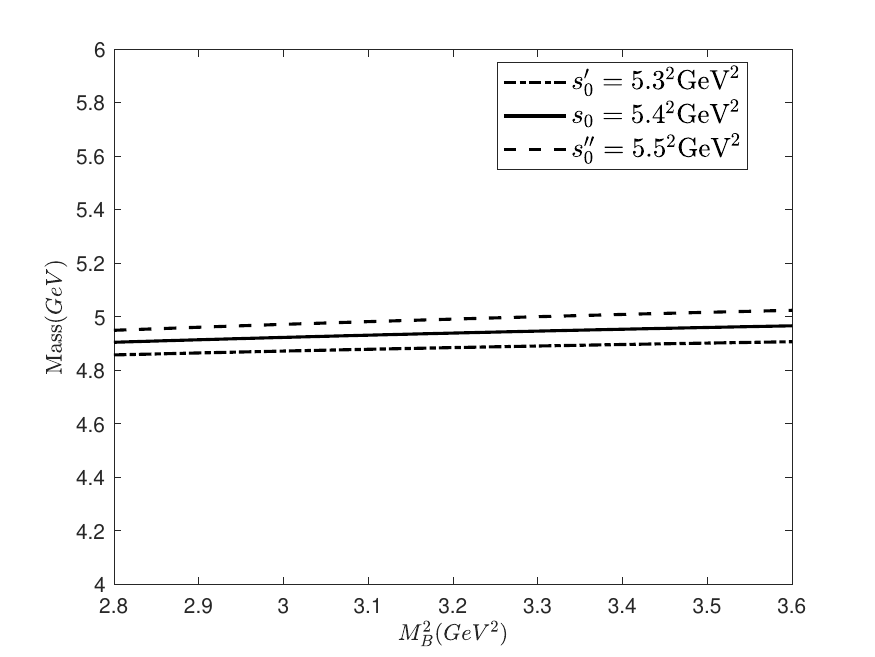}}
	\subfigure[]{
		\includegraphics[width=4.8cm]{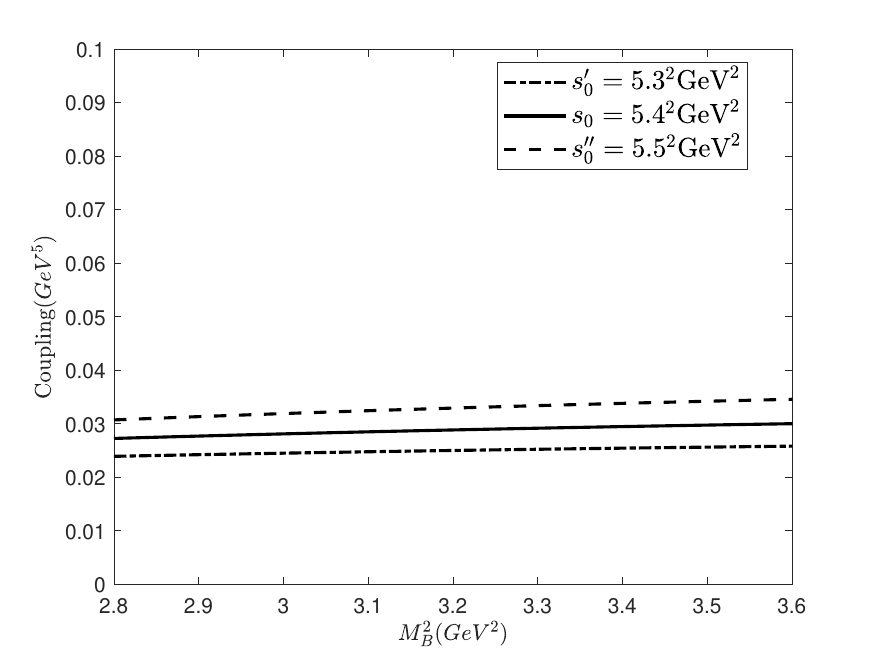}}
	\caption{(a) denotes the various ratios defined in (\ref{ratio}) as functions of $M^{2}_{B}$ with $s_{0}=(5.4~\mbox{GeV})^{2}$; (b) and (c) represent the mass and the current coupling constant varying with $M^{2}_{B}$ within the determined range at three different $s_{0}$ values showed in the figures, respectively.}\label{molecular}
\end{figure}

In Fig.\ref{molecular}(a), we show the various ratios defined in (\ref{ratio}) as functions of $M^{2}_{B}$ with $s_{0}=(5.4~\mbox{GeV})^{2}$. According to the criteria outlined above, we finally find the range of the Borel mass $M^{2}_{B}$, $M^{2}_{B}\in[2.8,3.6]~\mbox{GeV}^{2}$. From the Fig.\ref{molecular}(a), we also know that besides the quark condensate $\langle\bar{s}s\rangle$, the most important vacuum condensate is the mixed condensate $\langle g_{s}\bar{s}\sigma Gs\rangle$. This is because the Feynman diagrams corresponding to it are one order lower than those corresponding to the gluon condensate. Fig.\ref{molecular}(b) and (c) represent the mass and the current coupling constant varying with $M^{2}_{B}$ within the determined range at $s^{\prime}_{0}=(5.3~\mbox{GeV})^{2}$, $s_{0}=(5.4~\mbox{GeV})^{2}$, and $s^{\prime\prime}_{0}=(5.5~\mbox{GeV})^{2}$, respectively. According to the figures, we learn that the mass and the current coupling constant vary weakly with $M^{2}_{B}$ in $[2.8,3.6]~\mbox{GeV}^{2}$. The mass and the current coupling constant are $m_{M}=4.9392^{+0.0851}_{-0.0817}~\mbox{GeV}$ and $\lambda_{M}=2.8857^{+0.5729}_{-0.4928}\times10^{-2}~\mbox{GeV}^{5}$.

\begin{figure}[htb]
	\subfigure[]{
		\includegraphics[width=4.8cm]{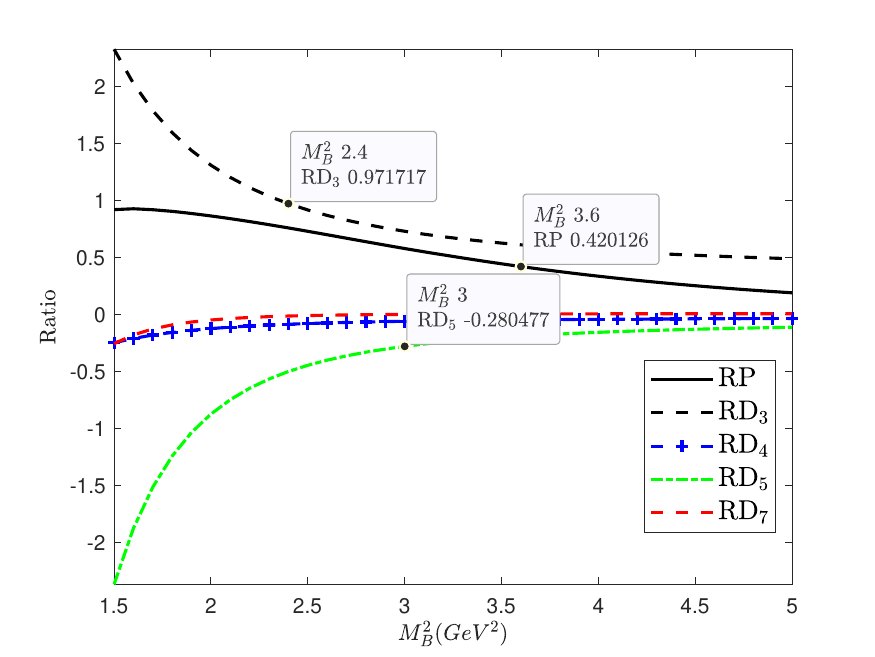}}
	\subfigure[]{
		\includegraphics[width=4.8cm]{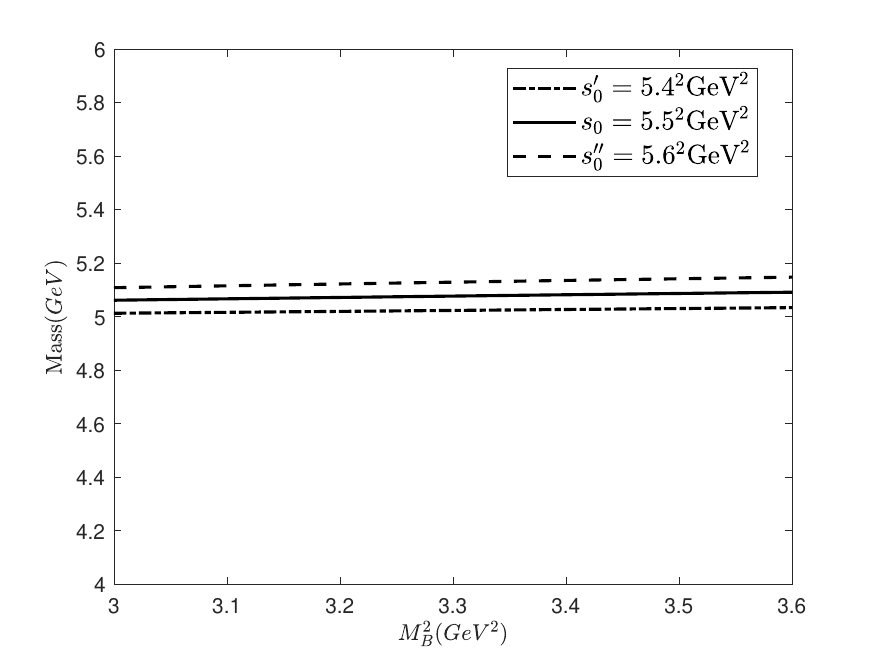}}
	\subfigure[]{
		\includegraphics[width=4.8cm]{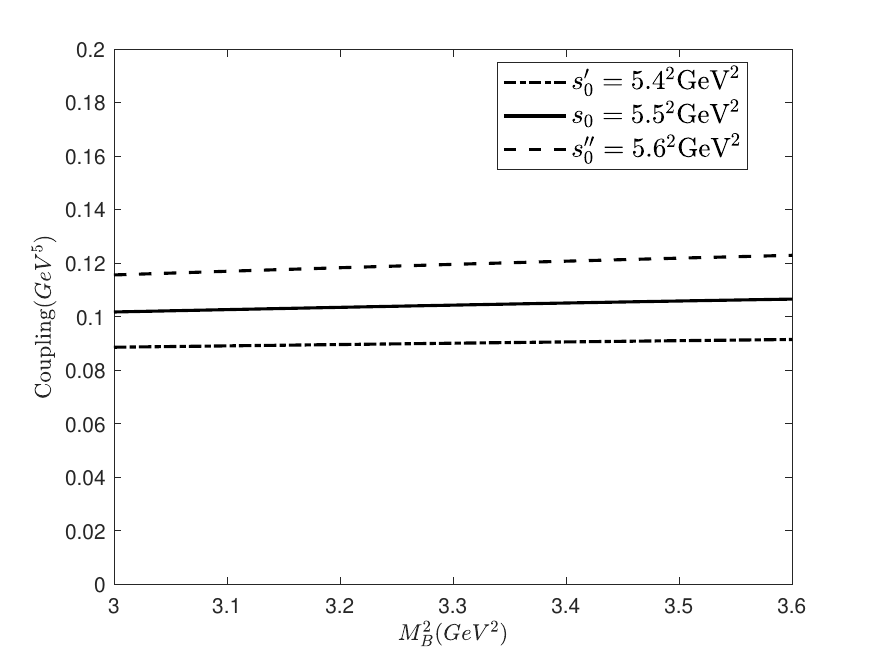}}
	\caption{(a) denotes the various ratios defined in (\ref{ratio}) as functions of $M^{2}_{B}$ with $s_{0}=(5.5~\mbox{GeV})^{2}$; (b) and (c) represent the mass and the current coupling constant varying with $M^{2}_{B}$ within the determined range at three different $s_{0}$ values showed in the figures, respectively.}\label{tetraquark1}
\end{figure}

\begin{figure}[htb]
	\subfigure[]{
		\includegraphics[width=4.8cm]{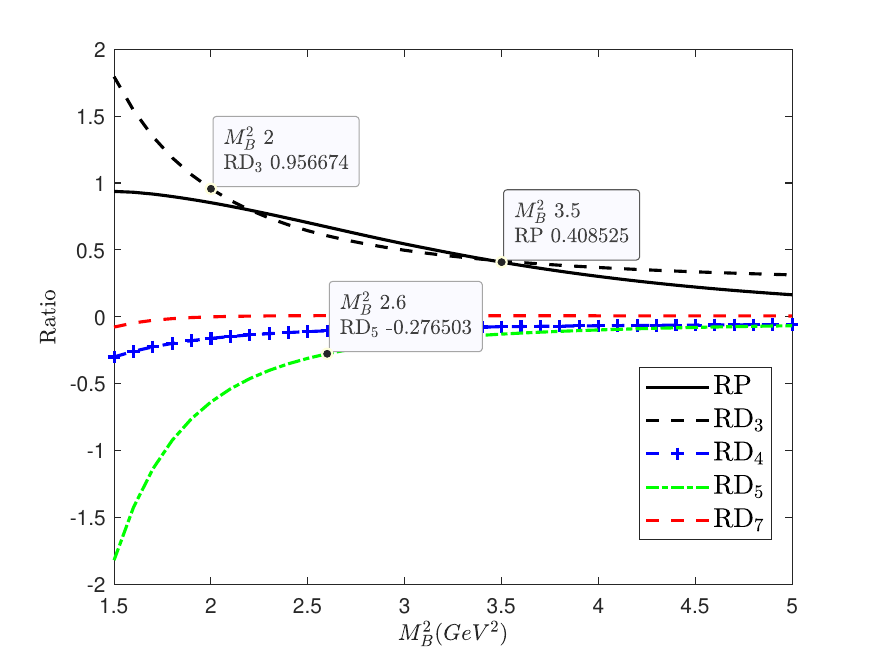}}
	\subfigure[]{
		\includegraphics[width=4.8cm]{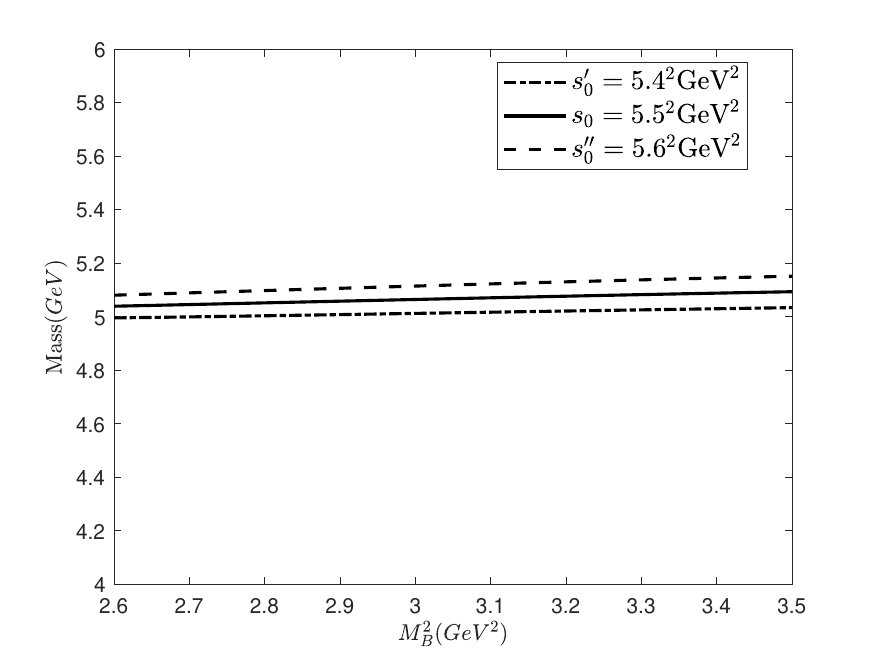}}
	\subfigure[]{
		\includegraphics[width=4.8cm]{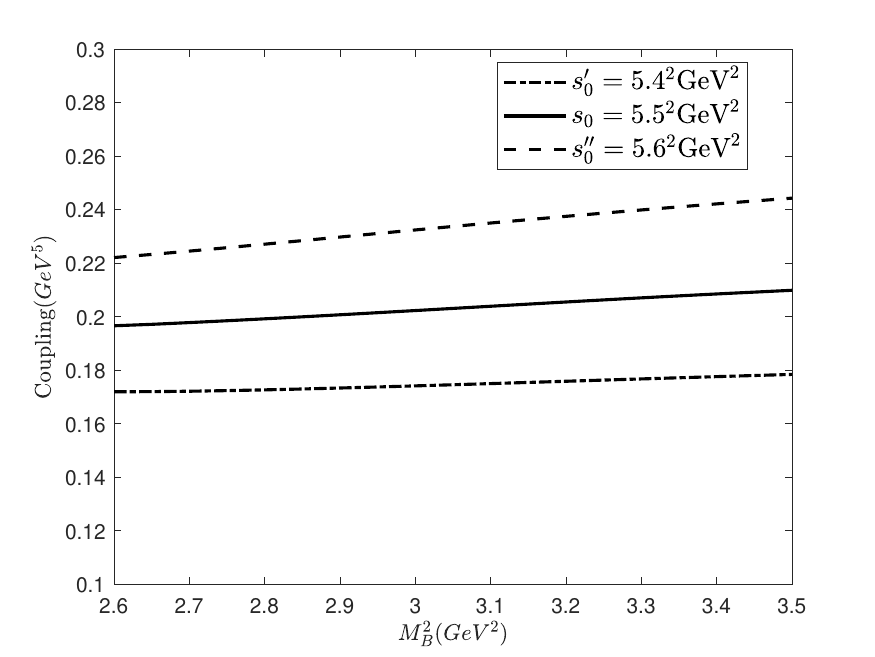}}
	\caption{(a) denotes the various ratios defined in (\ref{ratio}) as functions of $M^{2}_{B}$ with $s_{0}=(5.5~\mbox{GeV})^{2}$; (b) and (c) represent the mass and the current coupling constant varying with $M^{2}_{B}$ within the determined range at three different $s_{0}$ values showed in the figures, respectively.}\label{tetraquark2}
\end{figure}

The same analyses can be done for the sum rules of the $[cc]_{A}[\bar{c}\bar{s}]_{A}$ compact tetraquark state and the $[cc]_{T}[\bar{c}\bar{s}]_{T}$ compact tetraquark state. The results are shown in Figs.\ref{tetraquark1} and \ref{tetraquark2}. From Figs.\ref{tetraquark1}(a) and \ref{tetraquark2}(b), the allowed ranges of the Borel parameter $M^{2}_{B}$ are $[3.0,3.6]~\mbox{GeV}^{2}$ and $[2.6,3.5]~\mbox{GeV}^{2}$ for the $[cc]_{A}[\bar{c}\bar{s}]_{A}$ compact tetraquark state and the $[cc]_{T}[\bar{c}\bar{s}]_{T}$ compact tetraquark state, respectively. Figs.\ref{tetraquark1}(b) and (c) show the weak dependence on $M^{2}_{B}$ within $[3.0,3.6]~\mbox{GeV}^{2}$ of the mass and the current coupling constant of the $[cc]_{A}[\bar{c}\bar{s}]_{A}$ compact tetraquark state. The numerical values are $m_{T_{1}}=5.0774^{+0.0708}_{-0.0641}~\mbox{GeV}$ and $\lambda_{T_{1}}=1.0436^{+0.1862}_{-0.1573}\times10^{-1}~\mbox{GeV}^{5}$. From Figs.\ref{tetraquark2}(b) and (c), it is known that the mass and the current coupling constant of the $[cc]_{T}[\bar{c}\bar{s}]_{T}$ compact tetraquark state vary weakly with the Borel mass $M^{2}_{B}$ in $[2.6,3.5]~\mbox{GeV}^{2}$ and the results are $m_{T_{2}}=5.0679^{+0.0839}_{-0.0721}~\mbox{GeV}$, and $\lambda_{T_{2}}=2.0316^{+0.4119}_{-0.3119}\times10^{-1}~\mbox{GeV}^{5}$.

We summarize our results in Table \ref{results}.
\begin{table}[htb]
	\caption{The masses and the pole residues of the scalar triple-Heavy tetraquark states with quark content $cc\bar{c}\bar{s}$.(SP and IC stand for Spectrum Parameter and Interpolating Current, respectively.)}\label{results}
	\begin{tabular}{|c|c|c|c|}
		\hline
		\diagbox[width=3.8cm]{SP}{IC} & $J^{M}$ & $J^{T_{1}}$ & $J^{T_{2}}$\\
		\hline
		Mass(GeV) & $4.9392^{+0.0851}_{-0.0817}$ & $5.0774^{+0.0708}_{-0.0641}$ & $5.0679^{+0.0839}_{-0.0721}$ \\
		\hline
		Pole Residue($\mbox{GeV}^{5}$) & $2.8857^{+0.5729}_{-0.4928}\times10^{-2}$ & $1.0436^{+0.1862}_{-0.1573}\times10^{-1}$ & $2.0316^{+0.4119}_{-0.3119}\times10^{-1}$\\
		\hline
	\end{tabular}
\end{table}

\section{Conclusion}\label{sec4}
In this paper, the scalar triple-charm tetraquark states with strangeness are studied. We consider two types of internal quark configurations, $\eta_{c}D_{s}$ molecular state and diquark-antidiquark compact tetraquark states. For the molecular configuration, the interpolating current is $J^{M}(x)$. While for the diquark-antidiquark configuration, we use the interpolating currents of the following type, $J(x)=\epsilon_{ijn}\epsilon_{kln}[c^{T}_{i}(x)C\Gamma c_{j}(x)][\bar{c}_{k}C\Gamma\bar{s}^{T}_{l}(x)]$. Due to the symmetrical properties of $C\Gamma$, there are two possibles, $\gamma_{\mu}$ and $\sigma_{\mu\nu}$, and the corresponding interpolating currents are $J^{T_{1}}(x)$ and $J^{T_{2}}(x)$. With these interpolating currents, we derive the desired sum rules. The numerical values are $m_{M}=4.9392^{+0.0851}_{-0.0817}~\mbox{GeV}$, $\lambda_{M}=2.8857^{+0.5729}_{-0.4928}\times10^{-2}~\mbox{GeV}^{5}$, $m_{T_{1}}=5.0774^{+0.0708}_{-0.0641}~\mbox{GeV}$, $\lambda_{T_{1}}=1.0436^{+0.1862}_{-0.1573}\times10^{-1}~\mbox{GeV}^{5}$, $m_{T_{2}}=5.0679^{+0.0839}_{-0.0721}~\mbox{GeV}$, and $\lambda_{T_{2}}=2.0316^{+0.4119}_{-0.3119}\times10^{-1}~\mbox{GeV}^{5}$. Our investigation indicates that there is possible of scalar bound states with quark content $cc\bar{c}\bar{s}$, $\eta_{c}D_{s}$ molecular state and diquark-antidiquark compact tetraquark states.

\appendix
\section{The spectral density}\label{spectral_density}

On the theoretical side, we carry out the OPE up to dimension 7 for the spectral densities $\rho(s)$. The explicit expressions of the spectral densities are given below:
\begin{equation}
	\rho(s)=\rho_{0}+\rho_{3}(s)+\rho_{4}(s)+\rho_{5}(s)+\rho_{7}(s),
\end{equation}
with
\begin{eqnarray}
	\rho_{0}(s)=&&-\frac{m^{2}_{c}}{2^{11}\pi^{6}}\int^{a_{max}}_{a_{min}}da\int^{b_{max}}_{b_{min}}db\int^{1-a-b}_{c_{min}}dc\frac{a+b-11c}{a^{3}b^{3}c^{3}}\Delta^{3}\nonumber\\&&+\frac{33m^{3}_{c}m_{s}}{2^{11}\pi^{6}}\int^{a_{max}}_{a_{min}}da\int^{b_{max}}_{b_{min}}db\int^{1-a-b}_{c_{min}}dc\frac{\Delta^{2}}{a^{2}b^{2}c^{2}}\nonumber\\&&+\frac{11}{2^{13}\pi^{6}}\int^{a_{max}}_{a_{min}}da\int^{b_{max}}_{b_{min}}db\frac{\Delta^{\prime4}}{a^{3}b^{3}(1-a-b)^{3}}\nonumber\\&&-\frac{m_{c}m_{s}}{2^{11}\pi^{6}}\int^{a_{max}}_{a_{min}}da\int^{b_{max}}_{b_{min}}db\frac{(a+b)(1-a-b)-11ab}{a^{3}b^{3}(1-a-b)^{3}}\Delta^{\prime3},
\end{eqnarray}
\begin{eqnarray}
	\rho_{3}(s)=&&-\frac{11m^{3}_{c}\langle\bar{s}s\rangle}{2^{8}\pi^{4}}\int^{a_{max}}_{a_{min}}da\int^{b_{max}}_{b_{min}}db\frac{\Delta^{\prime}}{ab(1-a-b)}\nonumber\\&&+\frac{m_{c}\langle\bar{s}s\rangle}{2^{8}\pi^{4}}\int^{a_{max}}_{a_{min}}da\int^{b_{max}}_{b_{min}}db\frac{(a+b)(1-a-b)-11ab}{a^{2}b^{2}(1-a-b)^{2}}\Delta^{\prime}[\Delta^{\prime}+ab(1-a-b)s]\nonumber\\&&-\frac{m^{2}_{c}m_{s}\langle\bar{s}s\rangle}{2^{9}\pi^{4}}\int^{a_{max}}_{a_{min}}da\int^{b_{max}}_{b_{min}}db\frac{12(a+b)-11}{ab(1-a-b)}[2\Delta^{\prime}+ab(1-a-b)s]\nonumber\\&&+\frac{11m_{s}\langle\bar{s}s\rangle}{2^{9}\pi^{4}}\int^{a_{max}}_{a_{min}}da\int^{b_{max}}_{b_{min}}db\frac{1}{ab(1-a-b)}\{10\Delta^{\prime2}\nonumber\\&&+8\Delta^{\prime}m^{2}_{c}[(a+b)(1-a-b)+ab]+m^{4}_{c}[(a+b)(1-a-b)+ab]^{2}\},
\end{eqnarray}
\begin{eqnarray}
	\rho_{4}(s)=&&-\frac{m^{2}_{c}\langle g^{2}_{s}GG\rangle}{2^{13}\pi^{6}}\int^{a_{max}}_{a_{min}}da\int^{b_{max}}_{b_{min}}db\int^{1-a-b}_{c_{min}}dc\frac{1}{a^{3}b^{3}c^{3}}\nonumber\\&&[(a^{3}+b^{3})c^{2}+a^{2}b^{2}(a+b)-11(a^{2}+b^{2})c^{3}]\Delta\nonumber\\&&+\frac{m^{4}_{c}\langle g^{2}_{s}GG\rangle}{3\cdot2^{13}\pi^{6}}\int^{a_{max}}_{a_{min}}da\int^{b_{max}}_{b_{min}}db\int^{1-a-b}_{c_{min}}dc\frac{1}{a^{3}b^{3}c^{3}}\nonumber\\&&(a+b-11c)[a^{3}b^{3}+a^{3}c^{3}+b^{3}c^{3}]\nonumber\\&&+\frac{11m^{3}_{c}m_{s}\langle g^{2}_{s}GG\rangle}{2^{13}\pi^{6}}\int^{a_{max}}_{a_{min}}da\int^{b_{max}}_{b_{min}}db\int^{1-a-b}_{c_{min}}dc\frac{a^{2}b^{2}+a^{2}c^{2}+b^{2}c^{2}}{a^{2}b^{2}c^{2}}\nonumber\\&&-\frac{m^{2}_{c}\langle g^{2}_{s}GG\rangle}{2^{12}\pi^{6}}\int^{a_{max}}_{a_{min}}da\int^{b_{max}}_{b_{min}}db\int^{1-a-b}_{c_{min}}dc\frac{a^{2}+b^{2}-12c^{2}}{a^{2}b^{2}c^{2}}\Delta\nonumber\\&&-\frac{m^{2}_{c}\langle g^{2}_{s}GG\rangle}{2^{12}\pi^{6}}\int^{a_{max}}_{a_{min}}da\int^{b_{max}}_{b_{min}}db\int^{1-a-b}_{c_{min}}dc\frac{ab+ac+bc}{a^{2}b^{2}c^{2}}\Delta\nonumber\\&&-\frac{m^{3}_{c}m_{s}\langle g^{2}_{s}GG\rangle}{2^{12}\pi^{6}}\int^{a_{max}}_{a_{min}}da\int^{b_{max}}_{b_{min}}db\int^{1-a-b}_{c_{min}}dc\frac{a+b-11c}{abc}\nonumber\\&&-\frac{11m^{5}_{c}m_{s}\langle g^{2}_{s}GG\rangle}{3\cdot2^{13}\pi^{6}}\int^{a_{max}}_{a_{min}}da\int^{b_{max}}_{b_{min}}db\frac{a^{3}b^{3}+a^{3}c^{3}+b^{3}c^{3}}{a^{2}b^{2}c^{2}|abs-m^{2}_{c}(a+b)|}|_{c=c_{min}}\nonumber\\&&-\frac{11m^{2}_{c}\langle g^{2}_{s}GG\rangle}{3\cdot2^{13}\pi^{6}}\int^{a_{max}}_{a_{min}}da\int^{b_{max}}_{b_{min}}db\frac{a^{3}b^{3}+(a^{3}+b^{3})(1-a-b)^{3}}{a^{3}b^{3}(1-a-b)^{3}}\Delta^{\prime}\nonumber\\&&-\frac{m_{c}m_{s}\langle g^{2}_{s}GG\rangle}{2^{13}\pi^{6}}\int^{a_{max}}_{a_{min}}da\int^{b_{max}}_{b_{min}}db\frac{(a^{3}+b^{3})(1-a-b)^{3}-11a^{3}b^{3}}{a^{3}b^{3}(1-a-b)^{3}}\Delta^{\prime}\nonumber\\&&+\frac{m^{3}_{c}m_{s}\langle g^{2}_{s}GG\rangle}{3\cdot2^{13}\pi^{6}}\int^{a_{max}}_{a_{min}}da\int^{b_{max}}_{b_{min}}db\frac{1}{a^{3}b^{3}(1-a-b)^{3}}\nonumber\\&&[a^{3}b^{3}+(a^{3}+b^{3})(1-a-b)^{3}][(a+b)(1-a-b)-11ab]\nonumber\\&&-\frac{\langle g^{2}_{s}GG\rangle}{2^{14}\pi^{6}}\int^{a_{max}}_{a_{min}}da\int^{b_{max}}_{b_{min}}db\frac{12(a+b)-11}{a^{2}b^{2}(1-a-b)^{2}}\Delta^{\prime2}\nonumber\\&&+\frac{9m_{c}m_{s}\langle g^{2}_{s}GG\rangle}{2^{13}\pi^{6}}\int^{a_{max}}_{a_{min}}da\int^{b_{max}}_{b_{min}}db\frac{1}{ab(1-a-b)}\Delta^{\prime}\nonumber\\&&-\frac{m_{c}m_{s}\langle g^{2}_{s}GG\rangle}{2^{13}\pi^{6}}\int^{a_{max}}_{a_{min}}da\int^{b_{max}}_{b_{min}}db\frac{1}{a^{2}b^{2}(1-a-b)^{2}}\nonumber\\&&[(a+b)(1-a-b)^{2}+(1-a)a^{2}+(1-b)b^{2}]\Delta^{\prime}\nonumber\\&&+\frac{9m^{2}_{c}\langle g^{2}_{s}GG\rangle}{2^{13}\pi^{6}}\int^{a_{max}}_{a_{min}}da\int^{b_{max}}_{b_{min}}db\frac{1}{ab(1-a-b)}\Delta^{\prime}\nonumber\\&&-\frac{\langle g^{2}_{s}GG\rangle}{2^{13}\pi^{6}}\int^{a_{max}}_{a_{min}}da\int^{b_{max}}_{b_{min}}db\frac{(a+b)(1-a-b)-11ab}{a^{2}b^{2}(1-a-b)^{2}}\nonumber\\&&\Delta^{\prime}[\Delta^{\prime}+ab(1-a-b)s]\nonumber\\&&-\frac{m^{2}_{c}\langle g^{2}_{s}GG\rangle}{2^{13}\pi^{6}}\int^{a_{max}}_{a_{min}}da\int^{b_{max}}_{b_{min}}db\frac{1}{a^{2}b^{2}(1-a-b)^{2}}\nonumber\\&&[(a+b)(1-a-b)^{2}+(1-a)a^{2}+(1-b)b^{2}]\Delta^{\prime},
\end{eqnarray}
\begin{eqnarray}
	\rho_{5}(s)=&&\frac{11m^{3}_{c}\langle g_{s}\bar{s}\sigma Gs\rangle}{2^{9}\pi^{4}}\int^{a_{max}}_{a_{min}}da\int^{b_{max}}_{b_{min}}db\nonumber\\&&-\frac{3m_{c}\langle g_{s}\bar{s}\sigma Gs\rangle}{2^{9}\pi^{4}}\int^{a_{max}}_{a_{min}}da\int^{b_{max}}_{b_{min}}db\frac{(a+b)(1-a-b)-11ab}{ab(1-a-b)}\Delta^{\prime}\nonumber\\&&-\frac{3m_{c}\langle g_{s}\bar{s}\sigma Gs\rangle}{2^{9}\pi^{4}}\int^{a_{max}}_{a_{min}}da\int^{b_{max}}_{b_{min}}db[(a+b)(1-a-b)-11ab]s\nonumber\\&&+\frac{m^{2}_{c}m_{s}\langle g_{s}\bar{s}\sigma Gs\rangle}{2^{9}\pi^{4}}\int^{a_{max}}_{a_{min}}da\int^{b_{max}}_{b_{min}}db[12(a+b)-11]\nonumber\\&&-\frac{33m_{s}\langle g_{s}\bar{s}\sigma Gs\rangle}{2^{8}\pi^{4}}\int^{a_{max}}_{a_{min}}da\int^{b_{max}}_{b_{min}}dbab(1-a-b)s\nonumber\\&&-\frac{11m_{s}\langle g_{s}\bar{s}\sigma Gs\rangle}{2^{7}\pi^{4}}\int^{a_{max}}_{a_{min}}da\int^{b_{max}}_{b_{min}}db\Delta^{\prime}\nonumber\\&&+\frac{m^{3}_{c}\langle g_{s}\bar{s}\sigma Gs\rangle}{2^{9}\pi^{4}}\int^{a_{max}}_{a_{min}}da\int^{b_{max}}_{b_{min}}db\frac{(a+b)(1-a-b)-11ab}{ab(1-a-b)}\nonumber\\&&+\frac{m_{c}\langle g_{s}\bar{s}\sigma Gs\rangle}{2^{9}\pi^{4}}\int^{a_{max}}_{a_{min}}da\int^{b_{max}}_{b_{min}}db\frac{(a^{2}+b^{2})(1-a-b)^{2}-12a^{2}b^{2}}{ab(1-a-b)}s\nonumber\\&&+\frac{m_{c}\langle g_{s}\bar{s}\sigma Gs\rangle}{2^{9}\pi^{4}}\int^{a_{max}}_{a_{min}}da\int^{b_{max}}_{b_{min}}dbs\nonumber\\&&+\frac{m_{c}\langle g_{s}\bar{s}\sigma Gs\rangle}{2^{8}\pi^{4}}\int^{a_{max}}_{a_{min}}da\int^{b_{max}}_{b_{min}}db\frac{(a^{2}+b^{2})(1-a-b)^{2}-12a^{2}b^{2}}{a^{2}b^{2}(1-a-b)^{2}}\Delta^{\prime}\nonumber\\&&+\frac{m_{c}\langle g_{s}\bar{s}\sigma Gs\rangle}{2^{8}\pi^{4}}\int^{a_{max}}_{a_{min}}da\int^{b_{max}}_{b_{min}}db\frac{1}{ab(1-a-b)}\Delta^{\prime}\nonumber\\&&+\frac{11m^{3}_{c}\langle g_{s}\bar{s}\sigma Gs\rangle}{2^{10}\pi^{4}}\int^{1}_{0}da\int^{1-a}_{0}dbab(1-a-b)s\delta(\Delta^{\prime})\nonumber\\&&-\frac{m_{c}\langle g_{s}\bar{s}\sigma Gs\rangle}{2^{10}\pi^{4}}\int^{1}_{0}da\int^{1-a}_{0}db[(a+b)(1-a-b)-11ab]ab(1-a-b)s^{2}\delta(\Delta^{\prime})\nonumber\\&&+\frac{m^{2}_{c}m_{s}\langle g_{s}\bar{s}\sigma Gs\rangle}{3\cdot2^{10}\pi^{4}M^{2}_{B}}\int^{1}_{0}da\int^{1-a}_{0}db[12(a+b)-11]ab(1-a-b)s^{2}\delta(\Delta^{\prime})\nonumber\\&&+\frac{m^{2}_{c}m_{s}\langle g_{s}\bar{s}\sigma Gs\rangle}{3\cdot2^{8}\pi^{4}}\int^{1}_{0}da\int^{1-a}_{0}db[12(a+b)-11]ab(1-a-b)s\delta(\Delta^{\prime})\nonumber\\&&-\frac{11m_{s}\langle g_{s}\bar{s}\sigma Gs\rangle}{3\cdot2^{10}\pi^{4}M^{2}_{B}}\int^{1}_{0}da\int^{1-a}_{0}dba^{2}b^{2}(1-a-b)^{2}s^{3}\delta(\Delta^{\prime})\nonumber\\&&-\frac{33m_{s}\langle g_{s}\bar{s}\sigma Gs\rangle}{2^{10}\pi^{4}}\int^{1}_{0}da\int^{1-a}_{0}dba^{2}b^{2}(1-a-b)^{2}s^{2}\delta(\Delta^{\prime}),
\end{eqnarray}
\begin{eqnarray}
	\rho_{7}(s)=&&\frac{m_{c}\langle\bar{s}s\rangle\langle g^{2}_{s}GG\rangle}{3\cdot2^{10}\pi^{4}}\int^{a_{max}}_{a_{min}}da\int^{b_{max}}_{b_{min}}db\frac{(a^{3}+b^{3})(1-a-b)^{3}-11a^{3}b^{3}}{a^{2}b^{2}(1-a-b)^{2}}\nonumber\\&&-\frac{m_{c}\langle\bar{s}s\rangle\langle g^{2}_{s}GG\rangle}{2^{8}\pi^{4}}\int^{a_{max}}_{a_{min}}da\int^{b_{max}}_{b_{min}}db\nonumber\\&&+\frac{m_{c}\langle\bar{s}s\rangle\langle g^{2}_{s}GG\rangle}{3\cdot2^{10}\pi^{4}}\int^{a_{max}}_{a_{min}}da\int^{b_{max}}_{b_{min}}db\frac{(a+b)(1-a-b)+ab}{ab(1-a-b)}\nonumber\\&&-\frac{m_{s}\langle\bar{s}s\rangle\langle g^{2}_{s}GG\rangle}{2^{11}\pi^{4}}\int^{a_{max}}_{a_{min}}da\int^{b_{max}}_{b_{min}}db[12(a+b)-11]\nonumber\\&&-\frac{11m^{3}_{c}\langle\bar{s}s\rangle\langle g^{2}_{s}GG\rangle}{3\cdot2^{11}\pi^{4}}\int^{1}_{0}da\int^{1-a}_{0}db\frac{a^{2}b^{2}+(a^{2}+b^{2})(1-a-b)^{2}}{ab(1-a-b)}\delta(\Delta^{\prime})\nonumber\\&&+\frac{11m^{5}_{c}\langle\bar{s}s\rangle\langle g^{2}_{s}GG\rangle}{9\cdot2^{11}\pi^{4}M^{2}_{B}}\int^{1}_{0}da\int^{1-a}_{0}db\frac{a^{3}b^{3}+(a^{3}+b^{3})(1-a-b)^{3}}{a^{2}b^{2}(1-a-b)^{2}}\delta(\Delta^{\prime})\nonumber\\&&+\frac{m_{c}\langle\bar{s}s\rangle\langle g^{2}_{s}GG\rangle}{3\cdot2^{11}\pi^{4}}\int^{1}_{0}da\int^{1-a}_{0}db\frac{(a^{3}+b^{3})(1-a-b)^{3}-11a^{3}b^{3}}{ab(1-a-b)}s\delta(\Delta^{\prime})\nonumber\\&&-\frac{m^{3}_{c}\langle\bar{s}s\rangle\langle g^{2}_{s}GG\rangle}{9\cdot2^{11}\pi^{4}}\int^{1}_{0}da\int^{1-a}_{0}db\frac{1}{a^{2}b^{2}(1-a-b)^{2}}\nonumber\\&&[a^{3}b^{3}+(a^{3}+b^{3})(1-a-b)^{3}][(a+b)(1-a-b)-11ab]\delta(\Delta^{\prime})\nonumber\\&&-\frac{m^{3}_{c}\langle\bar{s}s\rangle\langle g^{2}_{s}GG\rangle}{9\cdot2^{11}\pi^{4}M^{2}_{B}}\int^{1}_{0}da\int^{1-a}_{0}db\frac{1}{a^{2}b^{2}(1-a-b)^{2}}\nonumber\\&&[a^{3}b^{3}+(a^{3}+b^{3})(1-a-b)^{3}][(a+b)(1-a-b)-11ab]s\delta(\Delta^{\prime})\nonumber\\&&-\frac{m^{2}_{c}m_{s}\langle\bar{s}s\rangle\langle g^{2}_{s}GG\rangle}{3\cdot2^{12}\pi^{4}}\int^{1}_{0}da\int^{1-a}_{0}db\frac{1}{ab(1-a-b)}\nonumber\\&&[(a^{3}+b^{3})(1-a-b)^{2}+a^{2}b^{2}(a+b)-11(a^{2}+b^{2})(1-a-b)^{3}]\delta(\Delta^{\prime})\nonumber\\&&-\frac{m^{2}_{c}m_{s}\langle\bar{s}s\rangle\langle g^{2}_{s}GG\rangle}{3\cdot2^{12}\pi^{4}M^{2}_{B}}\int^{1}_{0}da\int^{1-a}_{0}db\frac{1}{ab(1-a-b)}\nonumber\\&&[(a^{3}+b^{3})(1-a-b)^{2}+a^{2}b^{2}(a+b)-11(a^{2}+b^{2})(1-a-b)^{3}]s\delta(\Delta^{\prime})\nonumber\\&&+\frac{m^{6}_{c}m_{s}\langle\bar{s}s\rangle\langle g^{2}_{s}GG\rangle}{9\cdot2^{12}\pi^{4}M^{4}_{B}}\int^{1}_{0}da\int^{1-a}_{0}db\frac{[(a+b)(1-a-b)+ab]}{a^{3}b^{3}(1-a-b)^{3}}\nonumber\\&&[(a^{4}+b^{4})(1-a-b)^{3}+a^{3}b^{3}(a+b)-11(a^{3}+b^{3})(1-a-b)^{4}]\delta(\Delta^{\prime})\nonumber\\&&+\frac{m^{6}_{c}m_{s}\langle\bar{s}s\rangle\langle g^{2}_{s}GG\rangle}{9\cdot2^{12}\pi^{4}M^{4}_{B}}\int^{1}_{0}da\int^{1-a}_{0}db\frac{1}{a^{2}b^{2}(1-a-b)^{2}}\nonumber\\&&[(a^{2}+b^{2})(1-a-b)^{2}-11a^{2}b^{2}][(a+b)(1-a-b)+ab]\delta(\Delta^{\prime})\nonumber\\&&-\frac{11m^{2}_{c}m_{s}\langle\bar{s}s\rangle\langle g^{2}_{s}GG\rangle}{9\cdot2^{11}\pi^{4}}\int^{1}_{0}da\int^{1-a}_{0}db\frac{a^{3}b^{3}+(a^{3}+b^{3})(1-a-b)^{3}}{ab(1-a-b)}\delta(\Delta^{\prime})\nonumber\\&&-\frac{11m^{2}_{c}m_{s}\langle\bar{s}s\rangle\langle g^{2}_{s}GG\rangle}{9\cdot2^{12}\pi^{4}M^{4}_{B}}\int^{1}_{0}da\int^{1-a}_{0}db\frac{[a^{3}b^{3}+(a^{3}+b^{3})(1-a-b)^{3}]}{ab(1-a-b)}s^{2}\delta(\Delta^{\prime})\nonumber\\&&-\frac{11m^{2}_{c}m_{s}\langle\bar{s}s\rangle\langle g^{2}_{s}GG\rangle}{9\cdot2^{11}\pi^{4}M^{2}_{B}}\int^{1}_{0}da\int^{1-a}_{0}db\frac{[a^{3}b^{3}+(a^{3}+b^{3})(1-a-b)^{3}]}{ab(1-a-b)}s\delta(\Delta^{\prime})\nonumber\\&&-\frac{m^{3}_{c}\langle\bar{s}s\rangle\langle g^{2}_{s}GG\rangle}{3\cdot2^{10}\pi^{4}}\int^{1}_{0}da\int^{1-a}_{0}db[11-12(a+b)]\delta(\Delta^{\prime})\nonumber\\&&-\frac{m_{c}\langle\bar{s}s\rangle\langle g^{2}_{s}GG\rangle}{2^{9}\pi^{4}}\int^{1}_{0}da\int^{1-a}_{0}dbab(1-a-b)s\delta(\Delta^{\prime})\nonumber\\&&+\frac{m_{c}\langle\bar{s}s\rangle\langle g^{2}_{s}GG\rangle}{3\cdot2^{11}\pi^{4}}\int^{1}_{0}da\int^{1-a}_{0}db[(a+b)(1-a-b)+ab]s\delta(\Delta^{\prime})\nonumber\\&&-\frac{m^{2}_{c}m_{s}\langle\bar{s}s\rangle\langle g^{2}_{s}GG\rangle}{3\cdot2^{11}\pi^{4}}\int^{1}_{0}da\int^{1-a}_{0}db[(a+b)(1-a-b)+ab]\delta(\Delta^{\prime})\nonumber\\&&-\frac{m^{2}_{c}m_{s}\langle\bar{s}s\rangle\langle g^{2}_{s}GG\rangle}{3\cdot2^{11}\pi^{4}M^{2}_{B}}\int^{1}_{0}da\int^{1-a}_{0}db[(a+b)(1-a-b)+ab]s\delta(\Delta^{\prime})\nonumber\\&&-\frac{m^{2}_{c}m_{s}\langle\bar{s}s\rangle\langle g^{2}_{s}GG\rangle}{3\cdot2^{11}\pi^{4}}\int^{1}_{0}da\int^{1-a}_{0}db[a^{2}+b^{2}-12(1-a-b)^{2}]\delta(\Delta^{\prime})\nonumber\\&&-\frac{m^{2}_{c}m_{s}\langle\bar{s}s\rangle\langle g^{2}_{s}GG\rangle}{3\cdot2^{11}\pi^{4}M^{2}_{B}}\int^{1}_{0}da\int^{1-a}_{0}db[a^{2}+b^{2}-12(1-a-b)^{2}]s\delta(\Delta^{\prime})\nonumber\\&&+\frac{m_{s}\langle\bar{s}s\rangle\langle g^{2}_{s}GG\rangle}{3\cdot2^{10}\pi^{4}}\int^{1}_{0}da\int^{1-a}_{0}db[11-12(a+b)]ab(1-a-b)s\delta(\Delta^{\prime})\nonumber\\&&+\frac{m_{s}\langle\bar{s}s\rangle\langle g^{2}_{s}GG\rangle}{3\cdot2^{12}\pi^{4}M^{2}_{B}}\int^{1}_{0}da\int^{1-a}_{0}db[11-12(a+b)]ab(1-a-b)s^{2}\delta(\Delta^{\prime}),
\end{eqnarray}
for the interpolating current $J^{M}(x)$.

\begin{equation}
	\rho(s)=\rho_{0}+\rho_{3}(s)+\rho_{4}(s)+\rho_{5}(s)+\rho_{7}(s),
\end{equation}
with
\begin{eqnarray}
	\rho_{0}(s)=&&\frac{m^{2}_{c}}{2^{5}\pi^{6}}\int^{a_{max}}_{a_{min}}da\int^{b_{max}}_{b_{min}}db\int^{1-a-b}_{c_{min}}dc\frac{\Delta^{3}}{a^{2}b^{3}c^{3}}\nonumber\\&&+\frac{3m^{3}_{c}m_{s}}{2^{4}\pi^{6}}\int^{a_{max}}_{a_{min}}da\int^{b_{max}}_{b_{min}}db\int^{1-a-b}_{c_{min}}dc\frac{\Delta^{2}}{a^{2}b^{2}c^{2}}\nonumber\\&&+\frac{1}{2^{6}\pi^{6}}\int^{a_{max}}_{a_{min}}da\int^{b_{max}}_{b_{min}}db\frac{\Delta^{\prime4}}{a^{3}b^{3}(1-a-b)^{3}}\nonumber\\&&+\frac{m_{c}m_{s}}{2^{5}\pi^{6}}\int^{a_{max}}_{a_{min}}da\int^{b_{max}}_{b_{min}}db\frac{\Delta^{\prime3}}{a^{3}b^{2}(1-a-b)^{2}},
\end{eqnarray}
\begin{eqnarray}
	\rho_{3}(s)=&&-\frac{m^{3}_{c}\langle\bar{s}s\rangle}{2\pi^{4}}\int^{a_{max}}_{a_{min}}da\int^{b_{max}}_{b_{min}}db\frac{\Delta^{\prime}}{ab(1-a-b)}\nonumber\\&&-\frac{m_{c}\langle\bar{s}s\rangle}{2^{2}\pi^{4}}\int^{a_{max}}_{a_{min}}da\int^{b_{max}}_{b_{min}}db\frac{\Delta^{\prime}[\Delta^{\prime}+ab(1-a-b)s]}{a^{2}b(1-a-b)}\nonumber\\&&+\frac{m^{2}_{c}m_{s}\langle\bar{s}s\rangle}{2^{3}\pi^{4}}\int^{a_{max}}_{a_{min}}da\int^{b_{max}}_{b_{min}}db\frac{2\Delta^{\prime}+ab(1-a-b)s}{b(1-a-b)}\nonumber\\&&+\frac{m_{s}\langle\bar{s}s\rangle}{2^{2}\pi^{4}}\int^{a_{max}}_{a_{min}}da\int^{b_{max}}_{b_{min}}db\frac{1}{ab(1-a-b)}\{10\Delta^{\prime2}\nonumber\\&&+8\Delta^{\prime}m^{2}_{c}[(a+b)(1-a-b)+ab]+m^{4}_{c}[(a+b)(1-a-b)+ab]^{2}\},
\end{eqnarray}
\begin{eqnarray}
	\rho_{4}(s)=&&\frac{m^{2}_{c}\langle g^{2}_{s}GG\rangle}{2^{7}\pi^{6}}\int^{a_{max}}_{a_{min}}da\int^{b_{max}}_{b_{min}}db\int^{1-a-b}_{c_{min}}dc\frac{b^{2}+c^{2}}{b^{3}c^{3}}\Delta\nonumber\\&&-\frac{m^{4}_{c}\langle g^{2}_{s}GG\rangle}{3\cdot2^{7}\pi^{6}}\int^{a_{max}}_{a_{min}}da\int^{b_{max}}_{b_{min}}db\int^{1-a-b}_{c_{min}}dc\frac{a^{3}b^{3}+a^{3}c^{3}+b^{3}c^{3}}{a^{2}b^{3}c^{3}}\nonumber\\&&+\frac{m^{3}_{c}m_{s}\langle g^{2}_{s}GG\rangle}{2^{6}\pi^{6}}\int^{a_{max}}_{a_{min}}da\int^{b_{max}}_{b_{min}}db\int^{1-a-b}_{c_{min}}dc\frac{a^{2}b^{2}+a^{2}c^{2}+b^{2}c^{2}}{a^{2}b^{2}c^{2}}\nonumber\\&&+\frac{m^{2}_{c}\langle g^{2}_{s}GG\rangle}{2^{9}\pi^{6}}\int^{a_{max}}_{a_{min}}da\int^{b_{max}}_{b_{min}}db\int^{1-a-b}_{c_{min}}dc\frac{b+c}{ab^{2}c^{2}}\Delta\nonumber\\&&+\frac{m^{3}_{c}m_{s}\langle g^{2}_{s}GG\rangle}{2^{8}\pi^{6}}\int^{a_{max}}_{a_{min}}da\int^{b_{max}}_{b_{min}}db\int^{1-a-b}_{c_{min}}dc\frac{b+c}{abc}\nonumber\\&&-\frac{m^{5}_{c}m_{s}\langle g^{2}_{s}GG\rangle}{3\cdot2^{6}\pi^{6}}\int^{a_{max}}_{a_{min}}da\int^{b_{max}}_{b_{min}}db\frac{a^{3}b^{3}+a^{3}c^{3}+b^{3}c^{3}}{a^{2}b^{2}c^{2}|abs-m^{2}_{c}(a+b)|}|_{c=c_{min}}\nonumber\\&&-\frac{m^{2}_{c}\langle g^{2}_{s}GG\rangle}{3\cdot2^{6}\pi^{6}}\int^{a_{max}}_{a_{min}}da\int^{b_{max}}_{b_{min}}db\frac{a^{3}b^{3}+(a^{3}+b^{3})(1-a-b)^{3}}{a^{3}b^{3}(1-a-b)^{3}}\Delta^{\prime}\nonumber\\&&-\frac{m^{3}_{c}m_{s}\langle g^{2}_{s}GG\rangle}{3\cdot2^{7}\pi^{6}}\int^{a_{max}}_{a_{min}}da\int^{b_{max}}_{b_{min}}db\frac{a^{3}b^{3}+(a^{3}+b^{3})(1-a-b)^{3}}{a^{3}b^{2}(1-a-b)^{2}}\nonumber\\&&+\frac{m_{c}m_{s}\langle g^{2}_{s}GG\rangle}{2^{7}\pi^{6}}\int^{a_{max}}_{a_{min}}da\int^{b_{max}}_{b_{min}}db\frac{1}{a^{3}}\Delta^{\prime}\nonumber\\&&+\frac{\langle g^{2}_{s}GG\rangle}{2^{10}\pi^{6}}\int^{a_{max}}_{a_{min}}da\int^{b_{max}}_{b_{min}}db\frac{1-a}{a^{2}b^{2}(1-a-b)^{2}}\Delta^{\prime2}\nonumber\\&&+\frac{m_{c}m_{s}\langle g^{2}_{s}GG\rangle}{2^{9}\pi^{6}}\int^{a_{max}}_{a_{min}}da\int^{b_{max}}_{b_{min}}db\frac{1-3a}{a^{2}b(1-a-b)}\Delta^{\prime}\nonumber\\&&+\frac{\langle g^{2}_{s}GG\rangle}{2^{9}\pi^{6}}\int^{a_{max}}_{a_{min}}da\int^{b_{max}}_{b_{min}}db\frac{1-a}{ab^{2}(1-a-b)^{2}}\Delta^{\prime}[\Delta^{\prime}+ab(1-a-b)s]\nonumber\\&&+\frac{m^{2}_{c}\langle g^{2}_{s}GG\rangle}{2^{9}\pi^{6}}\int^{a_{max}}_{a_{min}}da\int^{b_{max}}_{b_{min}}db\frac{1-a}{b^{2}(1-a-b)^{2}}\Delta^{\prime}\nonumber\\&&-\frac{m^{2}_{c}\langle g^{2}_{s}GG\rangle}{2^{8}\pi^{6}}\int^{a_{max}}_{a_{min}}da\int^{b_{max}}_{b_{min}}db\frac{1}{ab(1-a-b)}\Delta^{\prime},
\end{eqnarray}
\begin{eqnarray}
	\rho_{5}(s)=&&\frac{m^{3}_{c}\langle g_{s}\bar{s}\sigma Gs\rangle}{2^{2}\pi^{4}}\int^{a_{max}}_{a_{min}}da\int^{b_{max}}_{b_{min}}db\nonumber\\&&+\frac{3m_{c}\langle g_{s}\bar{s}\sigma Gs\rangle}{2^{3}\pi^{4}}\int^{a_{max}}_{a_{min}}da\int^{b_{max}}_{b_{min}}db\frac{1}{a}[\Delta^{\prime}+ab(1-a-b)s]\nonumber\\&&-\frac{m^{2}_{c}m_{s}\langle g_{s}\bar{s}\sigma Gs\rangle}{2^{3}\pi^{4}}\int^{a_{max}}_{a_{min}}da\int^{b_{max}}_{b_{min}}dba\nonumber\\&&-\frac{m_{s}\langle g_{s}\bar{s}\sigma Gs\rangle}{2\pi^{4}}\int^{a_{max}}_{a_{min}}da\int^{b_{max}}_{b_{min}}db[2\Delta^{\prime}+3ab(1-a-b)s]\nonumber\\&&-\frac{m^{3}_{c}\langle g_{s}\bar{s}\sigma Gs\rangle}{2^{5}\pi^{4}}\int^{a_{max}}_{a_{min}}da\int^{b_{max}}_{b_{min}}db\frac{1-a}{b(1-a-b)}\nonumber\\&&-\frac{m_{c}\langle g_{s}\bar{s}\sigma Gs\rangle}{2^{6}\pi^{4}}\int^{a_{max}}_{a_{min}}da\int^{b_{max}}_{b_{min}}db\frac{1-a}{ab(1-a-b)}[2\Delta^{\prime}+ab(1-a-b)s]\nonumber\\&&+\frac{m^{3}_{c}\langle g_{s}\bar{s}\sigma Gs\rangle}{2^{3}\pi^{4}}\int^{1}_{0}da\int^{1-a}_{0}dbab(1-a-b)s\delta(\Delta^{\prime})\nonumber\\&&+\frac{m_{c}\langle g_{s}\bar{s}\sigma Gs\rangle}{2^{4}\pi^{4}}\int^{1}_{0}da\int^{1-a}_{0}dbab^{2}(1-a-b)^{2}s^{2}\delta(\Delta^{\prime})\nonumber\\&&-\frac{m^{2}_{c}m_{s}\langle g_{s}\bar{s}\sigma Gs\rangle}{3\cdot2^{4}\pi^{4}M^{2}_{B}}\int^{1}_{0}da\int^{1-a}_{0}dba^{2}b(1-a-b)s^{2}\delta(\Delta^{\prime})\nonumber\\&&-\frac{m^{2}_{c}m_{s}\langle g_{s}\bar{s}\sigma Gs\rangle}{3\cdot2^{2}\pi^{4}}\int^{1}_{0}da\int^{1-a}_{0}dba^{2}b(1-a-b)s\delta(\Delta^{\prime})\nonumber\\&&-\frac{m_{s}\langle g_{s}\bar{s}\sigma Gs\rangle}{3\cdot2^{3}\pi^{4}M^{2}_{B}}\int^{1}_{0}da\int^{1-a}_{0}dba^{2}b^{2}(1-a-b)^{2}s^{3}\delta(\Delta^{\prime})\nonumber\\&&-\frac{3m_{s}\langle g_{s}\bar{s}\sigma Gs\rangle}{2^{3}\pi^{4}}\int^{1}_{0}da\int^{1-a}_{0}dba^{2}b^{2}(1-a-b)^{2}s^{2}\delta(\Delta^{\prime}),
\end{eqnarray}
\begin{eqnarray}
	\rho_{7}(s)=&&-\frac{m_{c}\langle\bar{s}s\rangle\langle g^{2}_{s}GG\rangle}{3\cdot2^{4}\pi^{4}}\int^{a_{max}}_{a_{min}}da\int^{b_{max}}_{b_{min}}db\frac{b(1-a-b)}{a^{2}}\nonumber\\&&-\frac{m_{c}\langle\bar{s}s\rangle\langle g^{2}_{s}GG\rangle}{3\cdot2^{6}\pi^{4}}\int^{a_{max}}_{a_{min}}da\int^{b_{max}}_{b_{min}}db\frac{1-3a}{a}\nonumber\\&&+\frac{m_{s}\langle\bar{s}s\rangle\langle g^{2}_{s}GG\rangle}{2^{7}\pi^{4}}\int^{a_{max}}_{a_{min}}da\int^{b_{max}}_{b_{min}}db(1-a)\nonumber\\&&-\frac{m_{c}\langle\bar{s}s\rangle\langle g^{2}_{s}GG\rangle}{3\cdot2^{5}\pi^{4}}\int^{1}_{0}da\int^{1-a}_{0}db\frac{b^{2}(1-a-b)^{2}s}{a}\delta(\Delta^{\prime})\nonumber\\&&-\frac{m^{3}_{c}\langle\bar{s}s\rangle\langle g^{2}_{s}GG\rangle}{3\cdot2^{4}\pi^{4}}\int^{1}_{0}da\int^{1-a}_{0}db\frac{a^{2}b^{2}+(a^{2}+b^{2})(1-a-b)^{2}}{ab(1-a-b)}\delta(\Delta^{\prime})\nonumber\\&&+\frac{m^{5}_{c}\langle\bar{s}s\rangle\langle g^{2}_{s}GG\rangle}{9\cdot2^{4}\pi^{4}M^{2}_{B}}\int^{1}_{0}da\int^{1-a}_{0}db\frac{a^{3}b^{3}+(a^{3}+b^{3})(1-a-b)^{3}}{a^{2}b^{2}(1-a-b)^{2}}\delta(\Delta^{\prime})\nonumber\\&&+\frac{m^{3}_{c}\langle\bar{s}s\rangle\langle g^{2}_{s}GG\rangle}{9\cdot2^{5}\pi^{4}}\int^{1}_{0}da\int^{1-a}_{0}db\frac{a^{3}b^{3}+(a^{3}+b^{3})(1-a-b)^{3}}{a^{2}b(1-a-b)}\delta(\Delta^{\prime})\nonumber\\&&+\frac{m^{3}_{c}\langle\bar{s}s\rangle\langle g^{2}_{s}GG\rangle}{9\cdot2^{5}\pi^{4}M^{2}_{B}}\int^{1}_{0}da\int^{1-a}_{0}db\frac{[a^{3}b^{3}+(a^{3}+b^{3})(1-a-b)^{3}]s}{a^{2}b(1-a-b)}\delta(\Delta^{\prime})\nonumber\\&&+\frac{m^{2}_{c}m_{s}\langle\bar{s}s\rangle\langle g^{2}_{s}GG\rangle}{3\cdot2^{6}\pi^{4}}\int^{1}_{0}da\int^{1-a}_{0}db\frac{a^{2}[b^{2}+(1-a-b)^{2}]}{b(1-a-b)}\delta(\Delta^{\prime})\nonumber\\&&+\frac{m^{2}_{c}m_{s}\langle\bar{s}s\rangle\langle g^{2}_{s}GG\rangle}{3\cdot2^{6}\pi^{4}M^{2}_{B}}\int^{1}_{0}da\int^{1-a}_{0}db\frac{a^{2}[b^{2}+(1-a-b)^{2}]s}{b(1-a-b)}\delta(\Delta^{\prime})\nonumber\\&&-\frac{m^{4}_{c}m_{s}\langle\bar{s}s\rangle\langle g^{2}_{s}GG\rangle}{9\cdot2^{6}\pi^{4}M^{4}_{B}}\int^{1}_{0}da\int^{1-a}_{0}db\frac{[a^{3}b^{3}+(a^{3}+b^{3})(1-a-b)^{3}]s}{ab^{2}(1-a-b)^{2}}\delta(\Delta^{\prime})\nonumber\\&&-\frac{m^{2}_{c}m_{s}\langle\bar{s}s\rangle\langle g^{2}_{s}GG\rangle}{9\cdot2^{4}\pi^{4}}\int^{1}_{0}da\int^{1-a}_{0}db\frac{a^{3}b^{3}+(a^{3}+b^{3})(1-a-b)^{3}}{ab(1-a-b)}\delta(\Delta^{\prime})\nonumber\\&&-\frac{m^{2}_{c}m_{s}\langle\bar{s}s\rangle\langle g^{2}_{s}GG\rangle}{9\cdot2^{5}\pi^{4}M^{4}_{B}}\int^{1}_{0}da\int^{1-a}_{0}db\frac{[a^{3}b^{3}+(a^{3}+b^{3})(1-a-b)^{3}]s^{2}}{ab(1-a-b)}\delta(\Delta^{\prime})\nonumber\\&&-\frac{m^{2}_{c}m_{s}\langle\bar{s}s\rangle\langle g^{2}_{s}GG\rangle}{9\cdot2^{4}\pi^{4}M^{2}_{B}}\int^{1}_{0}da\int^{1-a}_{0}db\frac{[a^{3}b^{3}+(a^{3}+b^{3})(1-a-b)^{3}]s}{ab(1-a-b)}\delta(\Delta^{\prime})\nonumber\\&&-\frac{m^{3}_{c}\langle\bar{s}s\rangle\langle g^{2}_{s}GG\rangle}{3\cdot2^{6}\pi^{4}}\int^{1}_{0}da\int^{1-a}_{0}db(1-a)\delta(\Delta^{\prime})\nonumber\\&&-\frac{m_{c}\langle\bar{s}s\rangle\langle g^{2}_{s}GG\rangle}{3\cdot2^{7}\pi^{4}}\int^{1}_{0}da\int^{1-a}_{0}db(1-3a)b(1-a-b)s\delta(\Delta^{\prime})\nonumber\\&&+\frac{m^{2}_{c}m_{s}\langle\bar{s}s\rangle\langle g^{2}_{s}GG\rangle}{3\cdot2^{8}\pi^{4}}\int^{1}_{0}da\int^{1-a}_{0}dba(1-a)\delta(\Delta^{\prime})\nonumber\\&&+\frac{m^{2}_{c}m_{s}\langle\bar{s}s\rangle\langle g^{2}_{s}GG\rangle}{3\cdot2^{8}\pi^{4}M^{2}_{B}}\int^{1}_{0}da\int^{1-a}_{0}dba(1-a)s\delta(\Delta^{\prime})\nonumber\\&&+\frac{m_{s}\langle\bar{s}s\rangle\langle g^{2}_{s}GG\rangle}{3\cdot2^{6}\pi^{4}}\int^{1}_{0}da\int^{1-a}_{0}dba(1-a)b(1-a-b)s\delta(\Delta^{\prime})\nonumber\\&&+\frac{m_{s}\langle\bar{s}s\rangle\langle g^{2}_{s}GG\rangle}{3\cdot2^{8}\pi^{4}M^{2}_{B}}\int^{1}_{0}da\int^{1-a}_{0}dba(1-a)b(1-a-b)s^{2}\delta(\Delta^{\prime}),
\end{eqnarray}
for the interpolating current $J^{T_{1}}$.

\begin{equation}
	\rho(s)=\rho_{0}+\rho_{3}(s)+\rho_{4}(s)+\rho_{5}(s)+\rho_{7}(s),
\end{equation}
with
\begin{eqnarray}
	\rho_{0}(s)=&&\frac{9m^{3}_{c}m_{s}}{2^{3}\pi^{6}}\int^{a_{max}}_{a_{min}}da\int^{b_{max}}_{b_{min}}db\int^{1-a-b}_{c_{min}}dc\frac{\Delta^{2}}{a^{2}b^{2}c^{2}}\nonumber\\&&+\frac{3}{2^{5}\pi^{6}}\int^{a_{max}}_{a_{min}}da\int^{b_{max}}_{b_{min}}db\frac{\Delta^{\prime4}}{a^{3}b^{3}(1-a-b)^{3}},
\end{eqnarray}
\begin{eqnarray}
	\rho_{3}(s)=&&-\frac{3m^{3}_{c}\langle\bar{s}s\rangle}{\pi^{4}}\int^{a_{max}}_{a_{min}}da\int^{b_{max}}_{b_{min}}db\frac{\Delta^{\prime}}{ab(1-a-b)}\nonumber\\&&+\frac{3m_{s}\langle\bar{s}s\rangle}{2\pi^{4}}\int^{a_{max}}_{a_{min}}da\int^{b_{max}}_{b_{min}}db\frac{1}{ab(1-a-b)}\{10\Delta^{\prime2}\nonumber\\&&+8\Delta^{\prime}m^{2}_{c}[(a+b)(1-a-b)+ab]+m^{4}_{c}[(a+b)(1-a-b)+ab]^{2}\},
\end{eqnarray}
\begin{eqnarray}
	\rho_{4}(s)=&&\frac{3m^{3}_{c}m_{s}\langle g^{2}_{s}GG\rangle}{2^{5}\pi^{6}}\int^{a_{max}}_{a_{min}}da\int^{b_{max}}_{b_{min}}db\int^{1-a-b}_{c_{min}}dc\frac{a^{2}b^{2}+a^{2}c^{2}+b^{2}c^{2}}{a^{2}b^{2}c^{2}}\nonumber\\&&-\frac{m^{3}_{c}m_{s}\langle g^{2}_{s}GG\rangle}{2^{5}\pi^{6}}\int^{a_{max}}_{a_{min}}da\int^{b_{max}}_{b_{min}}db\int^{1-a-b}_{c_{min}}dc\frac{a-b-c}{abc}\nonumber\\&&-\frac{m^{5}_{c}m_{s}\langle g^{2}_{s}GG\rangle}{2^{5}\pi^{6}}\int^{a_{max}}_{a_{min}}da\int^{b_{max}}_{b_{min}}db\frac{a^{3}b^{3}+a^{3}c^{3}+b^{3}c^{3}}{a^{2}b^{2}c^{2}|abs-m^{2}_{c}(a+b)|}|_{c=c_{min}}\nonumber\\&&-\frac{m^{2}_{c}\langle g^{2}_{s}GG\rangle}{2^{5}\pi^{6}}\int^{a_{max}}_{a_{min}}da\int^{b_{max}}_{b_{min}}db\frac{a^{3}b^{3}+(a^{3}+b^{3})(1-a-b)^{3}}{a^{3}b^{3}(1-a-b)^{3}}\Delta^{\prime}\nonumber\\&&+\frac{\langle g^{2}_{s}GG\rangle}{2^{6}\pi^{6}}\int^{a_{max}}_{a_{min}}da\int^{b_{max}}_{b_{min}}db\frac{(a-b)(1-a-b)+ab}{a^{2}b^{2}(1-a-b)^{2}}\Delta^{\prime}[\Delta^{\prime}+ab(1-a-b)s]\nonumber\\&&+\frac{\langle g^{2}_{s}GG\rangle}{2^{7}\pi^{6}}\int^{a_{max}}_{a_{min}}da\int^{b_{max}}_{b_{min}}db\frac{1-2a}{a^{2}b^{2}(1-a-b)^{2}}\Delta^{\prime2},
\end{eqnarray}
\begin{eqnarray}
	\rho_{5}(s)=&&\frac{3m^{3}_{c}\langle g_{s}\bar{s}\sigma Gs\rangle}{2\pi^{4}}\int^{a_{max}}_{a_{min}}da\int^{b_{max}}_{b_{min}}db\nonumber\\&&-\frac{9m_{s}\langle g_{s}\bar{s}\sigma Gs\rangle}{\pi^{4}}\int^{a_{max}}_{a_{min}}da\int^{b_{max}}_{b_{min}}dbab(1-a-b)s\nonumber\\&&-\frac{6m_{s}\langle g_{s}\bar{s}\sigma Gs\rangle}{\pi^{4}}\int^{a_{max}}_{a_{min}}da\int^{b_{max}}_{b_{min}}db\Delta^{\prime}\nonumber\\&&-\frac{m^{3}_{c}\langle g_{s}\bar{s}\sigma Gs\rangle}{2^{2}\pi^{4}}\int^{a_{max}}_{a_{min}}da\int^{b_{max}}_{b_{min}}db\frac{(a-b)(1-a-b)+ab}{ab(1-a-b)}\nonumber\\&&+\frac{3m^{3}_{c}\langle g_{s}\bar{s}\sigma Gs\rangle}{2^{2}\pi^{4}}\int^{1}_{0}da\int^{1-a}_{0}dbab(1-a-b)s\delta(\Delta^{\prime})\nonumber\\&&-\frac{m_{s}\langle g_{s}\bar{s}\sigma Gs\rangle}{2^{2}\pi^{4}M^{2}_{B}}\int^{1}_{0}da\int^{1-a}_{0}dba^{2}b^{2}(1-a-b)^{2}s^{3}\delta(\Delta^{\prime})\nonumber\\&&-\frac{9m_{s}\langle g_{s}\bar{s}\sigma Gs\rangle}{2^{2}\pi^{4}}\int^{1}_{0}da\int^{1-a}_{0}dba^{2}b^{2}(1-a-b)^{2}s^{2}\delta(\Delta^{\prime}),
\end{eqnarray}
\begin{eqnarray}
	\rho_{7}(s)=&&\frac{m_{s}\langle\bar{s}s\rangle\langle g^{2}_{s}GG\rangle}{2^{4}\pi^{4}}\int^{a_{max}}_{a_{min}}da\int^{b_{max}}_{b_{min}}db(1-2a)\nonumber\\&&-\frac{m^{3}_{c}\langle\bar{s}s\rangle\langle g^{2}_{s}GG\rangle}{2^{3}\pi^{4}}\int^{1}_{0}da\int^{1-a}_{0}db\frac{a^{2}b^{2}+(a^{2}+b^{2})(1-a-b)^{2}}{ab(1-a-b)}\delta(\Delta^{\prime})\nonumber\\&&+\frac{m^{5}_{c}\langle\bar{s}s\rangle\langle g^{2}_{s}GG\rangle}{3\cdot2^{3}\pi^{4}M^{2}_{B}}\int^{1}_{0}da\int^{1-a}_{0}db\frac{a^{3}b^{3}+(a^{3}+b^{3})(1-a-b)^{3}}{a^{2}b^{2}(1-a-b)^{2}}\delta(\Delta^{\prime})\nonumber\\&&-\frac{m^{2}_{c}m_{s}\langle\bar{s}s\rangle\langle g^{2}_{s}GG\rangle}{3\cdot2^{3}\pi^{4}}\int^{1}_{0}da\int^{1-a}_{0}db\frac{a^{3}b^{3}+(a^{3}+b^{3})(1-a-b)^{3}}{ab(1-a-b)}\delta(\Delta^{\prime})\nonumber\\&&-\frac{m^{2}_{c}m_{s}\langle\bar{s}s\rangle\langle g^{2}_{s}GG\rangle}{3\cdot2^{4}\pi^{4}M^{4}_{B}}\int^{1}_{0}da\int^{1-a}_{0}db\frac{[a^{3}b^{3}+(a^{3}+b^{3})(1-a-b)^{3}]s^{2}}{ab(1-a-b)}\delta(\Delta^{\prime})\nonumber\\&&-\frac{m^{2}_{c}m_{s}\langle\bar{s}s\rangle\langle g^{2}_{s}GG\rangle}{3\cdot2^{3}\pi^{4}M^{2}_{B}}\int^{1}_{0}da\int^{1-a}_{0}db\frac{[a^{3}b^{3}+(a^{3}+b^{3})(1-a-b)^{3}]s}{ab(1-a-b)}\delta(\Delta^{\prime})\nonumber\\&&-\frac{m^{3}_{c}\langle\bar{s}s\rangle\langle g^{2}_{s}GG\rangle}{3\cdot2^{3}\pi^{4}}\int^{1}_{0}da\int^{1-a}_{0}db(1-2a)\delta(\Delta^{\prime})\nonumber\\&&+\frac{m_{s}\langle\bar{s}s\rangle\langle g^{2}_{s}GG\rangle}{3\cdot2^{3}\pi^{4}}\int^{1}_{0}da\int^{1-a}_{0}dba(1-2a)b(1-a-b)s\delta(\Delta^{\prime})\nonumber\\&&+\frac{m_{s}\langle\bar{s}s\rangle\langle g^{2}_{s}GG\rangle}{3\cdot2^{5}\pi^{4}M^{2}_{B}}\int^{1}_{0}da\int^{1-a}_{0}dba(1-2a)b(1-a-b)s^{2}\delta(\Delta^{\prime}),
\end{eqnarray}
for the interpolating current $J^{T_{2}}$.

In the above equations, $\Delta=abcs-(ab+ac+bc)m^{2}_{c}$, $\Delta^{\prime}=ab(1-a-b)s-[ab+a(1-a-b)+b(1-a-b)]m^{2}_{c}$, $a_{max}=\frac{s-3m^{2}_{c}+\sqrt{(s-3m^{2}_{c})^{2}-4sm^{2}_{c}}}{2s}$, $a_{min}=\frac{s-3m^{2}_{c}-\sqrt{(s-3m^{2}_{c})^{2}-4sm^{2}_{c}}}{2s}$, $b_{max}=\frac{1-a+\sqrt{(1-a)^{2}-a(1-a)m^{2}_{c}/(as-m^{2}_{c})}}{2}$, $b_{min}=\frac{1-a-\sqrt{(1-a)^{2}-a(1-a)m^{2}_{c}/(as-m^{2}_{c})}}{2}$, and $c_{min}=\frac{abm^{2}_{c}}{abs-(a+b)m^{2}_{c}}$.

\bibliographystyle{JHEP}
\bibliography{references}

\providecommand{\href}[2]{#2}\begingroup\raggedright\begin{thebibliography}{10}

\bibitem{ParticleDataGroup:2024cfk}
{\scshape Particle Data Group} collaboration, \emph{{Review of particle
  physics}}, \href{https://doi.org/10.1103/PhysRevD.110.030001}{\emph{Phys.
  Rev. D} {\bfseries 110} (2024) 030001}.

\bibitem{Belle:2003nnu}
{\scshape Belle} collaboration, \emph{{Observation of a narrow charmonium-like
  state in exclusive $B^\pm \to K^\pm \pi^+ \pi^- J/\psi$ decays}},
  \href{https://doi.org/10.1103/PhysRevLett.91.262001}{\emph{Phys. Rev. Lett.}
  {\bfseries 91} (2003) 262001}
  [\href{https://arxiv.org/abs/hep-ex/0309032}{{\ttfamily hep-ex/0309032}}].

\bibitem{Liu:2013waa}
X.~Liu, \emph{{An overview of $XYZ$ new particles}},
  \href{https://doi.org/10.1007/s11434-014-0407-2}{\emph{Chin. Sci. Bull.}
  {\bfseries 59} (2014) 3815}
  [\href{https://arxiv.org/abs/1312.7408}{{\ttfamily 1312.7408}}].

\bibitem{Hosaka:2016pey}
A.~Hosaka, T.~Iijima, K.~Miyabayashi, Y.~Sakai and S.~Yasui, \emph{{Exotic
  hadrons with heavy flavors: X, Y, Z, and related states}},
  \href{https://doi.org/10.1093/ptep/ptw045}{\emph{PTEP} {\bfseries 2016}
  (2016) 062C01} [\href{https://arxiv.org/abs/1603.09229}{{\ttfamily
  1603.09229}}].

\bibitem{Chen:2016qju}
H.-X.~Chen, W.~Chen, X.~Liu and S.-L.~Zhu, \emph{{The hidden-charm pentaquark
  and tetraquark states}},
  \href{https://doi.org/10.1016/j.physrep.2016.05.004}{\emph{Phys. Rept.}
  {\bfseries 639} (2016) 1} [\href{https://arxiv.org/abs/1601.02092}{{\ttfamily
  1601.02092}}].

\bibitem{Richard:2016eis}
J.-M.~Richard, \emph{{Exotic hadrons: review and perspectives}},
  \href{https://doi.org/10.1007/s00601-016-1159-0}{\emph{Few Body Syst.}
  {\bfseries 57} (2016) 1185}
  [\href{https://arxiv.org/abs/1606.08593}{{\ttfamily 1606.08593}}].

\bibitem{Lebed:2016hpi}
R.F.~Lebed, R.E.~Mitchell and E.S.~Swanson, \emph{{Heavy-Quark QCD Exotica}},
  \href{https://doi.org/10.1016/j.ppnp.2016.11.003}{\emph{Prog. Part. Nucl.
  Phys.} {\bfseries 93} (2017) 143}
  [\href{https://arxiv.org/abs/1610.04528}{{\ttfamily 1610.04528}}].

\bibitem{Guo:2017jvc}
F.-K.~Guo, C.~Hanhart, U.-G.~Mei\ss{}ner, Q.~Wang, Q.~Zhao and B.-S.~Zou,
  \emph{{Hadronic molecules}},
  \href{https://doi.org/10.1103/RevModPhys.90.015004}{\emph{Rev. Mod. Phys.}
  {\bfseries 90} (2018) 015004}
  [\href{https://arxiv.org/abs/1705.00141}{{\ttfamily 1705.00141}}].

\bibitem{Liu:2019zoy}
Y.-R.~Liu, H.-X.~Chen, W.~Chen, X.~Liu and S.-L.~Zhu, \emph{{Pentaquark and
  Tetraquark states}},
  \href{https://doi.org/10.1016/j.ppnp.2019.04.003}{\emph{Prog. Part. Nucl.
  Phys.} {\bfseries 107} (2019) 237}
  [\href{https://arxiv.org/abs/1903.11976}{{\ttfamily 1903.11976}}].

\bibitem{Brambilla:2019esw}
N.~Brambilla, S.~Eidelman, C.~Hanhart, A.~Nefediev, C.-P.~Shen, C.E.~Thomas
  et~al., \emph{{The $XYZ$ states: experimental and theoretical status and
  perspectives}},
  \href{https://doi.org/10.1016/j.physrep.2020.05.001}{\emph{Phys. Rept.}
  {\bfseries 873} (2020) 1} [\href{https://arxiv.org/abs/1907.07583}{{\ttfamily
  1907.07583}}].

\bibitem{Chen:2022asf}
H.-X.~Chen, W.~Chen, X.~Liu, Y.-R.~Liu and S.-L.~Zhu, \emph{{An updated review
  of the new hadron states}},
  \href{https://doi.org/10.1088/1361-6633/aca3b6}{\emph{Rept. Prog. Phys.}
  {\bfseries 86} (2023) 026201}
  [\href{https://arxiv.org/abs/2204.02649}{{\ttfamily 2204.02649}}].

\bibitem{Wang:2025sic}
Z.-G.~Wang, \emph{{Review of the QCD sum rules for exotic states}},
  \href{https://arxiv.org/abs/2502.11351}{{\ttfamily 2502.11351}}.

\bibitem{GomshiNobary:2003sf}
M.A.~Gomshi~Nobary, \emph{{Fragmentation production of Omega(ccc) and
  Omega(bbb) baryons}},
  \href{https://doi.org/10.1016/j.physletb.2002.12.001}{\emph{Phys. Lett. B}
  {\bfseries 559} (2003) 239}
  [\href{https://arxiv.org/abs/hep-ph/0408122}{{\ttfamily hep-ph/0408122}}].

\bibitem{GomshiNobary:2004mq}
M.A.~Gomshi~Nobary and R.~Sepahvand, \emph{{Fragmentation of triply heavy
  baryons}}, \href{https://doi.org/10.1103/PhysRevD.71.034024}{\emph{Phys. Rev.
  D} {\bfseries 71} (2005) 034024}
  [\href{https://arxiv.org/abs/hep-ph/0406148}{{\ttfamily hep-ph/0406148}}].

\bibitem{Brambilla:2005yk}
N.~Brambilla, A.~Vairo and T.~Rosch, \emph{{Effective field theory Lagrangians
  for baryons with two and three heavy quarks}},
  \href{https://doi.org/10.1103/PhysRevD.72.034021}{\emph{Phys. Rev. D}
  {\bfseries 72} (2005) 034021}
  [\href{https://arxiv.org/abs/hep-ph/0506065}{{\ttfamily hep-ph/0506065}}].

\bibitem{GomshiNobary:2005ur}
M.A.~Gomshi~Nobary and R.~Sepahvand, \emph{{An Ivestigation of triply heavy
  baryon production at hadron colliders}},
  \href{https://doi.org/10.1016/j.nuclphysb.2006.01.043}{\emph{Nucl. Phys. B}
  {\bfseries 741} (2006) 34}
  [\href{https://arxiv.org/abs/hep-ph/0508115}{{\ttfamily hep-ph/0508115}}].

\bibitem{Jia:2006gw}
Y.~Jia, \emph{{Variational study of weakly coupled triply heavy baryons}},
  \href{https://doi.org/10.1088/1126-6708/2006/10/073}{\emph{JHEP} {\bfseries
  10} (2006) 073} [\href{https://arxiv.org/abs/hep-ph/0607290}{{\ttfamily
  hep-ph/0607290}}].

\bibitem{GomshiNobary:2007ofo}
M.A.~Gomshi~Nobary, B.~Nikoobakht and J.~Naji, \emph{{Production of Omega(bbc)
  and Omega(bcc) baryons in quark diquark model}},
  \href{https://doi.org/10.1016/j.nuclphysa.2007.02.008}{\emph{Nucl. Phys. A}
  {\bfseries 789} (2007) 243}.

\bibitem{Martynenko:2007je}
A.P.~Martynenko, \emph{{Ground-state triply and doubly heavy baryons in a
  relativistic three-quark model}},
  \href{https://doi.org/10.1016/j.physletb.2008.04.030}{\emph{Phys. Lett. B}
  {\bfseries 663} (2008) 317}
  [\href{https://arxiv.org/abs/0708.2033}{{\ttfamily 0708.2033}}].

\bibitem{Patel:2008mv}
B.~Patel, A.~Majethiya and P.C.~Vinodkumar, \emph{{Masses and Magnetic moments
  of Triply Heavy Flavour Baryons in Hypercentral Model}},
  \href{https://doi.org/10.1007/s12043-009-0061-4}{\emph{Pramana} {\bfseries
  72} (2009) 679} [\href{https://arxiv.org/abs/0808.2880}{{\ttfamily
  0808.2880}}].

\bibitem{Meinel:2010pw}
S.~Meinel, \emph{{Prediction of the $Omega_{bbb}$ mass from lattice QCD}},
  \href{https://doi.org/10.1103/PhysRevD.82.114514}{\emph{Phys. Rev. D}
  {\bfseries 82} (2010) 114514}
  [\href{https://arxiv.org/abs/1008.3154}{{\ttfamily 1008.3154}}].

\bibitem{Chen:2011mb}
Y.-Q.~Chen and S.-Z.~Wu, \emph{{Production of Triply Heavy Baryons at LHC}},
  \href{https://doi.org/10.1007/JHEP08(2011)144}{\emph{JHEP} {\bfseries 08}
  (2011) 144} [\href{https://arxiv.org/abs/1106.0193}{{\ttfamily 1106.0193}}].

\bibitem{Flynn:2011gf}
J.M.~Flynn, E.~Hernandez and J.~Nieves, \emph{{Triply Heavy Baryons and Heavy
  Quark Spin Symmetry}},
  \href{https://doi.org/10.1103/PhysRevD.85.014012}{\emph{Phys. Rev. D}
  {\bfseries 85} (2012) 014012}
  [\href{https://arxiv.org/abs/1110.2962}{{\ttfamily 1110.2962}}].

\bibitem{Llanes-Estrada:2011gwu}
F.J.~Llanes-Estrada, O.I.~Pavlova and R.~Williams, \emph{{A First Estimate of
  Triply Heavy Baryon Masses from the pNRQCD Perturbative Static Potential}},
  \href{https://doi.org/10.1140/epjc/s10052-012-2019-9}{\emph{Eur. Phys. J. C}
  {\bfseries 72} (2012) 2019}
  [\href{https://arxiv.org/abs/1111.7087}{{\ttfamily 1111.7087}}].

\bibitem{Wang:2011ae}
Z.-G.~Wang, \emph{{Analysis of the Triply Heavy Baryon States with QCD Sum
  Rules}}, \href{https://doi.org/10.1088/0253-6102/58/5/17}{\emph{Commun.
  Theor. Phys.} {\bfseries 58} (2012) 723}
  [\href{https://arxiv.org/abs/1112.2274}{{\ttfamily 1112.2274}}].

\bibitem{Albertus:2012isp}
C.~Albertus, J.M.~Flynn, E.~Hernandez and J.~Nieves, \emph{{A nonrelativistic
  quark model evaluation of exclusive $b\to c$ semileptonic decay of triply
  heavy baryons and $c\to s,d$ semileptonic decay of $cb$ baryons}},
  \href{https://doi.org/10.22323/1.171.0146}{\emph{PoS} {\bfseries
  ConfinementX} (2012) 146} [\href{https://arxiv.org/abs/1301.3024}{{\ttfamily
  1301.3024}}].

\bibitem{Meinel:2012qz}
S.~Meinel, \emph{{Excited-state spectroscopy of triply-bottom baryons from
  lattice QCD}}, \href{https://doi.org/10.1103/PhysRevD.85.114510}{\emph{Phys.
  Rev. D} {\bfseries 85} (2012) 114510}
  [\href{https://arxiv.org/abs/1202.1312}{{\ttfamily 1202.1312}}].

\bibitem{Aliev:2012tt}
T.M.~Aliev, K.~Azizi and M.~Savci, \emph{{Masses and Residues of the Triply
  Heavy Spin-1/2 Baryons}},
  \href{https://doi.org/10.1007/JHEP04(2013)042}{\emph{JHEP} {\bfseries 04}
  (2013) 042} [\href{https://arxiv.org/abs/1212.6065}{{\ttfamily 1212.6065}}].

\bibitem{Padmanath:2013zfa}
M.~Padmanath, R.G.~Edwards, N.~Mathur and M.~Peardon, \emph{{Spectroscopy of
  triply-charmed baryons from lattice QCD}},
  \href{https://doi.org/10.1103/PhysRevD.90.074504}{\emph{Phys. Rev. D}
  {\bfseries 90} (2014) 074504}
  [\href{https://arxiv.org/abs/1307.7022}{{\ttfamily 1307.7022}}].

\bibitem{Aliev:2014lxa}
T.M.~Aliev, K.~Azizi and M.~Savc\i{}, \emph{{Properties of triply heavy
  spin-3/2 baryons}},
  \href{https://doi.org/10.1088/0954-3899/41/6/065003}{\emph{J. Phys. G}
  {\bfseries 41} (2014) 065003}
  [\href{https://arxiv.org/abs/1404.2091}{{\ttfamily 1404.2091}}].

\bibitem{Wei:2015gsa}
K.-W.~Wei, B.~Chen and X.-H.~Guo, \emph{{Masses of doubly and triply charmed
  baryons}}, \href{https://doi.org/10.1103/PhysRevD.92.076008}{\emph{Phys. Rev.
  D} {\bfseries 92} (2015) 076008}
  [\href{https://arxiv.org/abs/1503.05184}{{\ttfamily 1503.05184}}].

\bibitem{Wei:2016jyk}
K.-W.~Wei, B.~Chen, N.~Liu, Q.-Q.~Wang and X.-H.~Guo, \emph{{Spectroscopy of
  singly, doubly, and triply bottom baryons}},
  \href{https://doi.org/10.1103/PhysRevD.95.116005}{\emph{Phys. Rev. D}
  {\bfseries 95} (2017) 116005}
  [\href{https://arxiv.org/abs/1609.02512}{{\ttfamily 1609.02512}}].

\bibitem{Shah:2017jkr}
Z.~Shah and A.K.~Rai, \emph{{Masses and Regge trajectories of triply heavy
  $\Omega_{ccc}$ and $\Omega_{bbb}$ baryons}},
  \href{https://doi.org/10.1140/epja/i2017-12386-2}{\emph{Eur. Phys. J. A}
  {\bfseries 53} (2017) 195}.

\bibitem{Wang:2018utj}
W.~Wang and J.~Xu, \emph{{Weak Decays of Triply Heavy Baryons}},
  \href{https://doi.org/10.1103/PhysRevD.97.093007}{\emph{Phys. Rev. D}
  {\bfseries 97} (2018) 093007}
  [\href{https://arxiv.org/abs/1803.01476}{{\ttfamily 1803.01476}}].

\bibitem{Shah:2018div}
Z.~Shah and A.K.~Rai, \emph{{Ground and Excited State Masses of the $\varOmega
  _\textit{bbc}$ Baryon}},
  \href{https://doi.org/10.1007/s00601-018-1398-3}{\emph{Few Body Syst.}
  {\bfseries 59} (2018) 76}.

\bibitem{Yang:2019lsg}
G.~Yang, J.~Ping, P.G.~Ortega and J.~Segovia, \emph{{Triply heavy baryons in
  the constituent quark model}},
  \href{https://doi.org/10.1088/1674-1137/44/2/023102}{\emph{Chin. Phys. C}
  {\bfseries 44} (2020) 023102}
  [\href{https://arxiv.org/abs/1904.10166}{{\ttfamily 1904.10166}}].

\bibitem{Wang:2019gal}
Z.-G.~Wang, \emph{{Triply-charmed dibaryon states or two-baryon scattering
  states from QCD sum rules}},
  \href{https://doi.org/10.1103/PhysRevD.102.034008}{\emph{Phys. Rev. D}
  {\bfseries 102} (2020) 034008}
  [\href{https://arxiv.org/abs/1912.07230}{{\ttfamily 1912.07230}}].

\bibitem{Liu:2019vtx}
M.-S.~Liu, Q.-F.~L\"u and X.-H.~Zhong, \emph{{Triply charmed and bottom baryons
  in a constituent quark model}},
  \href{https://doi.org/10.1103/PhysRevD.101.074031}{\emph{Phys. Rev. D}
  {\bfseries 101} (2020) 074031}
  [\href{https://arxiv.org/abs/1912.11805}{{\ttfamily 1912.11805}}].

\bibitem{Alomayrah:2020qyw}
N.~Alomayrah and T.~Barakat, \emph{{The excited states of triply-heavy baryons
  in QCD sum rules}},
  \href{https://doi.org/10.1140/epja/s10050-020-00062-7}{\emph{Eur. Phys. J. A}
  {\bfseries 56} (2020) 76}.

\bibitem{Wang:2020avt}
Z.-G.~Wang, \emph{{Analysis of the triply-heavy baryon states with the QCD sum
  rules}}, \href{https://doi.org/10.1007/s43673-021-00006-3}{\emph{AAPPS Bull.}
  {\bfseries 31} (2021) 5} [\href{https://arxiv.org/abs/2010.08939}{{\ttfamily
  2010.08939}}].

\bibitem{Wu:2021tzo}
R.-H.~Wu, Y.-S.~Zuo, C.~Meng, Y.-Q.~Ma and K.-T.~Chao, \emph{{NLO effects for
  \ensuremath{\Omega} QQQ baryons in QCD Sum Rules}},
  \href{https://doi.org/10.1088/1674-1137/ac0b3c}{\emph{Chin. Phys. C}
  {\bfseries 45} (2021) 093103}
  [\href{https://arxiv.org/abs/2104.07384}{{\ttfamily 2104.07384}}].

\bibitem{Mutuk:2021zes}
H.~Mutuk and U.~\"Ozdem, \emph{{Magnetic moments of spin\textendash{}1/2 triply
  heavy baryons: a study of light-cone QCD and quark\textendash{}diquark
  model}}, \href{https://doi.org/10.1140/epjp/s13360-022-02724-5}{\emph{Eur.
  Phys. J. Plus} {\bfseries 137} (2022) 508}
  [\href{https://arxiv.org/abs/2107.04361}{{\ttfamily 2107.04361}}].

\bibitem{Huang:2021jxt}
F.~Huang, J.~Xu and X.-R.~Zhang, \emph{{Deciphering weak decays of triply heavy
  baryons by SU(3) analysis}},
  \href{https://doi.org/10.1140/epjc/s10052-021-09729-x}{\emph{Eur. Phys. J. C}
  {\bfseries 81} (2021) 976}
  [\href{https://arxiv.org/abs/2107.13958}{{\ttfamily 2107.13958}}].

\bibitem{Faustov:2021qqf}
R.N.~Faustov and V.O.~Galkin, \emph{{Triply heavy baryon spectroscopy in the
  relativistic quark model}},
  \href{https://doi.org/10.1103/PhysRevD.105.014013}{\emph{Phys. Rev. D}
  {\bfseries 105} (2022) 014013}
  [\href{https://arxiv.org/abs/2111.07702}{{\ttfamily 2111.07702}}].

\bibitem{Wang:2022ias}
W.~Wang and Z.-P.~Xing, \emph{{Weak decays of triply heavy baryons in light
  front approach}},
  \href{https://doi.org/10.1016/j.physletb.2022.137402}{\emph{Phys. Lett. B}
  {\bfseries 834} (2022) 137402}
  [\href{https://arxiv.org/abs/2203.14446}{{\ttfamily 2203.14446}}].

\bibitem{Zhao:2022vfr}
Z.-X.~Zhao, F.-W.~Zhang and Q.~Yang, \emph{{Weak decays of triply heavy
  baryons}}, \href{https://doi.org/10.1140/epjc/s10052-025-13780-3}{\emph{Eur.
  Phys. J. C} {\bfseries 85} (2025) 106}
  [\href{https://arxiv.org/abs/2204.00759}{{\ttfamily 2204.00759}}].

\bibitem{Li:2022vbc}
J.-B.~Li, L.-C.~Gui, W.~Qin, W.~Sun and J.~Liang, \emph{{Triply charmed baryons
  mass decomposition from lattice QCD*}},
  \href{https://doi.org/10.1088/1674-1137/adc0f4}{\emph{Chin. Phys. C}
  {\bfseries 49} (2025) 063103}
  [\href{https://arxiv.org/abs/2211.04713}{{\ttfamily 2211.04713}}].

\bibitem{Wu:2022fpj}
S.-Z.~Wu, P.~Wu and Y.-W.~Li, \emph{{Production of the triply heavy
  $\Omega_{ccc}$ and $\Omega_{bbb}$ baryons at $e^+e^-$ colliders}},
  \href{https://arxiv.org/abs/2211.17061}{{\ttfamily 2211.17061}}.

\bibitem{Zhao:2023imq}
Y.-C.~Zhao, C.-M.~Tang and L.~Tang, \emph{{Mass predictions of triply heavy
  hybrid baryons via QCD sum rules}},
  \href{https://doi.org/10.1140/epjc/s10052-023-11825-z}{\emph{Eur. Phys. J. C}
  {\bfseries 83} (2023) 654}
  [\href{https://arxiv.org/abs/2303.15173}{{\ttfamily 2303.15173}}].

\bibitem{Oudichhya:2023pkg}
J.~Oudichhya, K.~Gandhi and A.k.~Rai, \emph{{Investigation of $\Omega _{ccb}$
  and $\Omega _{cbb}$ baryons in Regge phenomenology}},
  \href{https://doi.org/10.1007/s12043-023-02630-0}{\emph{Pramana} {\bfseries
  97} (2023) 151} [\href{https://arxiv.org/abs/2304.05110}{{\ttfamily
  2304.05110}}].

\bibitem{Zhao:2023qww}
J.~Zhao and S.~Shi, \emph{{Triply heavy baryons QQQ in vacuum and in a hot QCD
  medium}}, \href{https://doi.org/10.1103/PhysRevC.109.024901}{\emph{Phys. Rev.
  C} {\bfseries 109} (2024) 024901}
  [\href{https://arxiv.org/abs/2311.04594}{{\ttfamily 2311.04594}}].

\bibitem{Najjar:2024deh}
Z.R.~Najjar, K.~Azizi and H.R.~Moshfegh, \emph{{Properties of the ground and
  excited states of triply heavy spin-1/2 baryons}},
  \href{https://doi.org/10.1140/epjc/s10052-024-12960-x}{\emph{Eur. Phys. J. C}
  {\bfseries 84} (2024) 612}
  [\href{https://arxiv.org/abs/2402.14348}{{\ttfamily 2402.14348}}].

\bibitem{deArenaza:2024dhe}
N.M.~de~Arenaza, J.J.~G\'alvez-Viruet and F.J.~Llanes-Estrada,
  \emph{{Triply-heavy/strange baryons with Cornell potential on a quantum
  computer}}, \href{https://doi.org/10.1140/epja/s10050-024-01430-3}{\emph{Eur.
  Phys. J. A} {\bfseries 60} (2024) 216}
  [\href{https://arxiv.org/abs/2407.07232}{{\ttfamily 2407.07232}}].

\bibitem{Xie:2024lfo}
J.-Q.~Xie, H.~Song and J.-K.~Chen, \emph{{Regge trajectories for the triply
  heavy bottom-charm baryons in the diquark picture}},
  \href{https://doi.org/10.1140/epjc/s10052-024-13438-6}{\emph{Eur. Phys. J. C}
  {\bfseries 84} (2024) 1048}
  [\href{https://arxiv.org/abs/2407.18280}{{\ttfamily 2407.18280}}].

\bibitem{Najjar:2024ngm}
Z.R.~Najjar, K.~Azizi and H.R.~Moshfegh, \emph{{Semileptonic decay of the
  triply heavy \ensuremath{\Omega}ccb to the observed \ensuremath{\Xi}cc++
  state}}, \href{https://doi.org/10.1103/PhysRevD.111.014016}{\emph{Phys. Rev.
  D} {\bfseries 111} (2025) 014016}
  [\href{https://arxiv.org/abs/2410.01602}{{\ttfamily 2410.01602}}].

\bibitem{Dhindsa:2024erk}
N.S.~Dhindsa, D.~Chakraborty, A.~Radhakrishnan, N.~Mathur and M.~Padmanath,
  \emph{{Precise study of triply charmed baryons ($\Omega_{ccc}$)}},
  \href{https://arxiv.org/abs/2411.12729}{{\ttfamily 2411.12729}}.

\bibitem{Yu:2025gdg}
G.-L.~Yu, Z.-Y.~Li, Z.-G.~Wang and Z.~Zhou, \emph{{Systematic analysis of the
  mass spectra of triply heavy baryons}},
  \href{https://doi.org/10.1140/epjc/s10052-025-14261-3}{\emph{Eur. Phys. J. C}
  {\bfseries 85} (2025) 543}
  [\href{https://arxiv.org/abs/2501.01803}{{\ttfamily 2501.01803}}].

\bibitem{Salehi:2025hjn}
N.~Salehi, \emph{{A novel approach for spectroscopic study of $\Omega_{bbc}$
  baryon in the hypercentral constituent Quark model}},
  \href{https://doi.org/10.1142/S0217732324502201}{\emph{Mod. Phys. Lett. A}
  {\bfseries 40} (2025) 2450220}.

\bibitem{Najjar:2025dzl}
Z.R.~Najjar and K.~Azizi, \emph{{Investigation of triply heavy spin-3/2 baryons
  in their ground and excited states}},
  \href{https://arxiv.org/abs/2504.06822}{{\ttfamily 2504.06822}}.

\bibitem{Chen:2016ont}
K.~Chen, X.~Liu, J.~Wu, Y.-R.~Liu and S.-L.~Zhu, \emph{{Triply heavy tetraquark
  states with the $QQ\bar{Q}\bar{q}$ configuration}},
  \href{https://doi.org/10.1140/epja/i2017-12199-3}{\emph{Eur. Phys. J. A}
  {\bfseries 53} (2017) 5} [\href{https://arxiv.org/abs/1609.06117}{{\ttfamily
  1609.06117}}].

\bibitem{Jiang:2017tdc}
J.-F.~Jiang, W.~Chen and S.-L.~Zhu, \emph{{Triply heavy $QQ\bar Q\bar q$
  tetraquark states}},
  \href{https://doi.org/10.1103/PhysRevD.96.094022}{\emph{Phys. Rev. D}
  {\bfseries 96} (2017) 094022}
  [\href{https://arxiv.org/abs/1708.00142}{{\ttfamily 1708.00142}}].

\bibitem{Liu:2019mxw}
Y.~Liu, M.A.~Nowak and I.~Zahed, \emph{{Heavy Holographic Exotics: Tetraquarks
  as Efimov States}},
  \href{https://doi.org/10.1103/PhysRevD.100.126023}{\emph{Phys. Rev. D}
  {\bfseries 100} (2019) 126023}
  [\href{https://arxiv.org/abs/1904.05189}{{\ttfamily 1904.05189}}].

\bibitem{Xing:2019wil}
Y.~Xing, \emph{{Weak decays of triply heavy tetraquarks
  ${b{\bar{c}}}{b{\bar{q}}}$}},
  \href{https://doi.org/10.1140/epjc/s10052-020-7625-3}{\emph{Eur. Phys. J. C}
  {\bfseries 80} (2020) 57} [\href{https://arxiv.org/abs/1910.11593}{{\ttfamily
  1910.11593}}].

\bibitem{Weng:2021ngd}
X.-Z.~Weng, W.-Z.~Deng and S.-L.~Zhu, \emph{{Triply heavy tetraquark states}},
  \href{https://doi.org/10.1103/PhysRevD.105.034026}{\emph{Phys. Rev. D}
  {\bfseries 105} (2022) 034026}
  [\href{https://arxiv.org/abs/2109.05243}{{\ttfamily 2109.05243}}].

\bibitem{Lu:2021kut}
Q.-F.~L\"u, D.-Y.~Chen, Y.-B.~Dong and E.~Santopinto, \emph{{Triply-heavy
  tetraquarks in an extended relativized quark model}},
  \href{https://doi.org/10.1103/PhysRevD.104.054026}{\emph{Phys. Rev. D}
  {\bfseries 104} (2021) 054026}
  [\href{https://arxiv.org/abs/2107.13930}{{\ttfamily 2107.13930}}].

\bibitem{Liu:2022jdl}
X.~Liu, Y.~Tan, D.~Chen, H.~Huang and J.~Ping, \emph{{Possible triply heavy
  tetraquark states in a chiral quark model}},
  \href{https://doi.org/10.1103/PhysRevD.107.054019}{\emph{Phys. Rev. D}
  {\bfseries 107} (2023) 054019}
  [\href{https://arxiv.org/abs/2205.08281}{{\ttfamily 2205.08281}}].

\bibitem{Mutuk:2023yev}
H.~Mutuk, \emph{{Flavor exotic triply-heavy tetraquark states in AdS/QCD
  potential}},
  \href{https://doi.org/10.1140/epjc/s10052-023-11526-7}{\emph{Eur. Phys. J. C}
  {\bfseries 83} (2023) 358}
  [\href{https://arxiv.org/abs/2305.03358}{{\ttfamily 2305.03358}}].

\bibitem{Zhu:2023lbx}
Z.-H.~Zhu, W.-X.~Zhang and D.~Jia, \emph{{Triply heavy tetraquark states:
  masses and other properties}},
  \href{https://doi.org/10.1140/epjc/s10052-024-12700-1}{\emph{Eur. Phys. J. C}
  {\bfseries 84} (2024) 344}
  [\href{https://arxiv.org/abs/2312.01908}{{\ttfamily 2312.01908}}].

\bibitem{Yang:2024nyc}
G.~Yang, J.~Ping and J.~Segovia, \emph{{Triply charm and bottom tetraquarks in
  a constituent quark model}},
  \href{https://doi.org/10.1103/PhysRevD.110.054036}{\emph{Phys. Rev. D}
  {\bfseries 110} (2024) 054036}
  [\href{https://arxiv.org/abs/2407.14548}{{\ttfamily 2407.14548}}].

\bibitem{Zhang:2024jvv}
W.-S.~Zhang and L.~Tang, \emph{{Investigating triply heavy tetraquark states
  through QCD sum rules}},  \href{https://arxiv.org/abs/2412.11531}{{\ttfamily
  2412.11531}}.

\bibitem{Li:2025fmf}
S.-Y.~Li, Y.-R.~Liu, Z.-L.~Man, C.-R.~Shu, Z.-G.~Si and J.~Wu, \emph{{Triply
  Heavy Tetraquark States in a Mass-Splitting Model}},
  \href{https://doi.org/10.3390/sym17020170}{\emph{Symmetry} {\bfseries 17}
  (2025) 170} [\href{https://arxiv.org/abs/2501.16105}{{\ttfamily
  2501.16105}}].

\bibitem{Galkin:2025ubt}
V.O.~Galkin and E.M.~Savchenko, \emph{{Masses of Ground States of Triply Heavy
  Tetraquarks}}, \href{https://doi.org/10.1134/S1063779624701594}{\emph{Phys.
  Part. Nucl.} {\bfseries 56} (2025) 330}.

\bibitem{Guo:2013xga}
F.-K.~Guo, C.~Hidalgo-Duque, J.~Nieves and M.P.~Valderrama,
  \emph{{Heavy-antiquark\textendash{}diquark symmetry and heavy hadron
  molecules: Are there triply heavy pentaquarks?}},
  \href{https://doi.org/10.1103/PhysRevD.88.054014}{\emph{Phys. Rev. D}
  {\bfseries 88} (2013) 054014}
  [\href{https://arxiv.org/abs/1305.4052}{{\ttfamily 1305.4052}}].

\bibitem{Chen:2017jjn}
R.~Chen, A.~Hosaka and X.~Liu, \emph{{Prediction of triple-charm molecular
  pentaquarks}}, \href{https://doi.org/10.1103/PhysRevD.96.114030}{\emph{Phys.
  Rev. D} {\bfseries 96} (2017) 114030}
  [\href{https://arxiv.org/abs/1711.09579}{{\ttfamily 1711.09579}}].

\bibitem{Wang:2018ihk}
Z.-G.~Wang, \emph{{Analysis of the triply-charmed pentaquark states with QCD
  sum rules}}, \href{https://doi.org/10.1140/epjc/s10052-018-5786-0}{\emph{Eur.
  Phys. J. C} {\bfseries 78} (2018) 300}
  [\href{https://arxiv.org/abs/1801.08419}{{\ttfamily 1801.08419}}].

\bibitem{Li:2018vhp}
S.-Y.~Li, Y.-R.~Liu, Y.-N.~Liu, Z.-G.~Si and J.~Wu, \emph{{Pentaquark states
  with the $QQQq\bar{q}$ configuration in a simple model}},
  \href{https://doi.org/10.1140/epjc/s10052-019-6589-7}{\emph{Eur. Phys. J. C}
  {\bfseries 79} (2019) 87} [\href{https://arxiv.org/abs/1809.08072}{{\ttfamily
  1809.08072}}].

\bibitem{Wang:2019aoc}
F.-L.~Wang, R.~Chen, Z.-W.~Liu and X.~Liu, \emph{{Possible triple-charm
  molecular pentaquarks from $\Xi_{cc}D_1/\Xi_{cc}D_2^*$ interactions}},
  \href{https://doi.org/10.1103/PhysRevD.99.054021}{\emph{Phys. Rev. D}
  {\bfseries 99} (2019) 054021}
  [\href{https://arxiv.org/abs/1901.01542}{{\ttfamily 1901.01542}}].

\bibitem{An:2019idk}
H.-T.~An, Q.-S.~Zhou, Z.-W.~Liu, Y.-R.~Liu and X.~Liu, \emph{{Exotic pentaquark
  states with the $qqQQ\bar{Q}$ configuration}},
  \href{https://doi.org/10.1103/PhysRevD.100.056004}{\emph{Phys. Rev. D}
  {\bfseries 100} (2019) 056004}
  [\href{https://arxiv.org/abs/1905.07858}{{\ttfamily 1905.07858}}].

\bibitem{Wang:2024yjp}
Z.-Y.~Wang, C.-W.~Xiao, Z.-F.~Sun and X.~Liu, \emph{{Possible molecules of
  triple-heavy pentaquarks within the extended local hidden gauge formalism}},
  \href{https://doi.org/10.1103/PhysRevD.110.076014}{\emph{Phys. Rev. D}
  {\bfseries 110} (2024) 076014}
  [\href{https://arxiv.org/abs/2407.13319}{{\ttfamily 2407.13319}}].

\bibitem{Shifman:1978bx}
M.A.~Shifman, A.I.~Vainshtein and V.I.~Zakharov, \emph{{QCD and Resonance
  Physics. Theoretical Foundations}},
  \href{https://doi.org/10.1016/0550-3213(79)90022-1}{\emph{Nucl. Phys. B}
  {\bfseries 147} (1979) 385}.

\bibitem{Shifman:1978by}
M.A.~Shifman, A.I.~Vainshtein and V.I.~Zakharov, \emph{{QCD and Resonance
  Physics: Applications}},
  \href{https://doi.org/10.1016/0550-3213(79)90023-3}{\emph{Nucl. Phys. B}
  {\bfseries 147} (1979) 448}.

\bibitem{Reinders:1984sr}
L.J.~Reinders, H.~Rubinstein and S.~Yazaki, \emph{{Hadron Properties from QCD
  Sum Rules}}, \href{https://doi.org/10.1016/0370-1573(85)90065-1}{\emph{Phys.
  Rept.} {\bfseries 127} (1985) 1}.

\bibitem{Colangelo:2000dp}
P.~Colangelo and A.~Khodjamirian, \emph{{QCD sum rules, a modern perspective}},
   \href{https://arxiv.org/abs/hep-ph/0010175}{{\ttfamily hep-ph/0010175}}.

\bibitem{Albuquerque:2018jkn}
R.M.~Albuquerque, J.M.~Dias, K.P.~Khemchandani, A.~Mart\'\i{}nez~Torres,
  F.S.~Navarra, M.~Nielsen et~al., \emph{{QCD sum rules approach to the $X,~Y$
  and $Z$ states}}, \href{https://doi.org/10.1088/1361-6471/ab2678}{\emph{J.
  Phys. G} {\bfseries 46} (2019) 093002}
  [\href{https://arxiv.org/abs/1812.08207}{{\ttfamily 1812.08207}}].

\end{thebibliography}\endgroup

\end{document}